\documentclass{article}
\usepackage[T1]{fontenc}
\usepackage{graphicx}
\usepackage{mathtools}
\usepackage{pgfplots}
\usepgfplotslibrary{groupplots}
\pgfplotsset{compat=1.18}
\definecolor{Aone}{HTML}{3C64D8}
\definecolor{Atwo}{HTML}{009A9A}
\definecolor{Athree}{HTML}{F76860}

\usepackage{amssymb,amsmath,amsthm,mathrsfs,enumitem}
\usepackage[margin=1.3in]{geometry}
\usepackage{booktabs,tabularx,array}
\usepackage{xcolor}
\usepackage{hyperref}
\definecolor{pink}{rgb}{1,0,1}

\def\beq{\begin{eqnarray}}
\def\eeq{\end{eqnarray}}
\newcommand{\nn}{\nonumber}
\newcommand{\cLstar}{\mathcal{L}^{*}}
\theoremstyle{plain}

\def\XXint#1#2#3{{\setbox0=\hbox{$#1{#2#3}{\int}$}%
\vcenter{\hbox{$#2#3$}}\kern-.5\wd0}}

\theoremstyle{definition}

\newcommand{\cG}{\mathcal{G}}
\newcommand{\cL}{\mathcal{L}}
\newcommand{\pstar}{\mathbf{p}^{*}}
\newcommand{\mca}{\operatorname{MCA}}
\newcommand{\keywords}[1]{{\renewcommand{\thefootnote}{}%
  \footnotetext{\emph{Keywords:} #1}}}

\title{The Shape of Individual Fitness Distributions and the Dynamics of Ecological Communities}
\author{Julie Rowlett \and Leo Tingqvist}
\begin{document}
\maketitle

\keywords{individual variability, game theory, stochastic differential equation, microbe ecology, ecological community.}
\begin{abstract}
Developing a mechanistic understanding of biogeography and biodiversity that predicts species abundances, distributions, and fluctuations is a central goal of community ecology. We develop a model, that can be seen as an extension of Hubbell's neutral theory, in which a  focal species competes with the aggregate of all other species in a local community of fixed size.  Each individual has its own competitive ability, and the shape of the distribution of these abilities is incorporated in the model, represented by distributions rather than single species-level values. Under complete mixing, symmetric competitive encounters without ties recover Hubbell’s neutral model. We further incorporate immigration from a larger metacommunity and derive a diffusion approximation for the focal species’ relative abundance.  The model shows that competitive ability distributions with the same mean and variance can nevertheless yield different patterns of local abundance. Thus, species that appear equivalent when summarized by trait statistics need not be ecologically equivalent. Our framework links individual-level variability to population-level persistence and abundance and provides a way to examine how competition and immigration jointly shape ecological outcomes.  
\end{abstract}

\section{Introduction}\label{s:intro}
Understanding why species differ in abundance, and why some persist while others disappear, is a central problem in community ecology. Local abundance reflects the combined effects of differences in ecological performance, demographic stochasticity, and the movement of individuals between communities \cite{Vellend2016}. These processes connect events involving individual organisms to changes in community composition. However, individuals belonging to the same species need not perform identically, and the consequences of their differences may depend on how variation is distributed within the population. Establishing how individual variability affects population dynamics therefore requires a connection between the distribution of individual characteristics and the demographic events through which species gain or lose abundance.

This problem is particularly relevant to microbial communities, where substantial diversity can occur among organisms competing for relatively few limiting resources. Hutchinson's paradox of the plankton drew attention to the difficulty of reconciling this diversity with competitive exclusion under equilibrium resource competition \cite{Hutchinson1961}. Subsequent work has shown that the outcome depends on the assumptions used to describe competition and community dynamics. For example, resource competition can generate oscillations and chaos that permit many species to persist on few resources \cite{HuismanWeissing1999}. Alongside such mechanisms, variation among individuals within a species raises a complementary question: how does the internal composition of a population influence its competitive relationship with the surrounding community?

Hubbell's neutral theory provides a useful demographic baseline for addressing this question \cite{Hubbell2001}. Under ecological equivalence, individuals of different species have the same per-capita demographic rates, while random births, deaths, and immigration generate changes in species abundance. Consequently, abundance can fluctuate and local extinction can occur even when no species has a systematic competitive advantage. Neutrality in this sense is an assumption about demographic performance; it does not require organisms to be identical in every trait. Conversely, equality in a mean trait value does not by itself establish demographic equivalence. This distinction motivates a model in which the relationship between individual characteristics and population-level competitive performance is derived explicitly, while retaining demographic stochasticity and immigration.

There is substantial biological motivation for considering distributions of individual characteristics. Phenotypic heterogeneity occurs even among genetically identical microorganisms exposed to similar environmental conditions, and can affect functions associated with growth and survival \cite{Ackermann2015}. At the community level, observations from phytoplankton assemblages show that diversity in traits measured on individual organisms can predict community properties better than species richness or evenness \cite{FontanaThomasMoldoveanu2018}. These observations do not identify a universal demographic effect of intraspecific variation, but they indicate that individual-level measurements contain ecological information that species abundances alone do not capture. The challenge is to determine when differences among individuals translate into differences in population dynamics. We address this by using distributions of competitive ability to calculate how often individuals win, lose, or tie in pairwise encounters. We then link those outcomes to replacement events that change local species abundance. 

In \cite{MendenDeuerRowlett2014,MendenDeuerRowlett2019,MendenDeuerRowlett2021,RowlettKarlssonNursultanov2022,KarlssonGerleeRowlett2025}, Rowlett et al. developed a model that interpolates between variable individuals within a species of microbes and the ramifications for the entire species. A key discovery is that species with identical mean competitive ability are not necessarily neutral towards each other in competition. Species with identical mean competitive ability but different distributions across variable individuals may be superior or inferior in competition with each other. This is biologically plausible because substantial phenotypic heterogeneity is ubiquitous in microbial populations, including variation among genetically identical individuals \cite{Ackermann2015}. Moreover, empirical studies have shown that intraspecific heterogeneity can directly affect population-level survival, growth, and competitive ability \cite{MendenDeuerRowlett2014,RyderheimKiorboe2024,HamerMatthiessenPulinaHattich2022,GallegoNarwani2022}. In natural yeast populations, for example, phenotypic heterogeneity in stress resistance was a stronger predictor of survival under environmental stress than mean resistance, and populations with greater heterogeneity exhibited a competitive advantage under stress \cite{Bishop2007,Holland2014}. Their experiments specifically tested survival under chemical and environmental stress and found that phenotypic heterogeneity within genetically uniform yeast populations can produce rare subpopulations that survive otherwise stressful conditions. Variation in multiple microbial growth traits can produce very nontrivial competitive outcomes, even with a single limiting resource \cite{ManhartShakhnovich2018}. Similarly, phenotypic diversification within isogenic microbial populations can enhance viability and metabolic activity under oligotrophic conditions \cite{Kundu2020}. These observations suggest that the distribution of individual phenotypes, rather than its mean alone, can be an important determinant of population success.

Based on these observations, in this study we represent each species by a distribution of individual competitive
abilities and calculate encounter outcomes by comparing individuals
sampled from competing distributions
\cite{MendenDeuerRowlett2014,RowlettKarlssonNursultanov2022}.
Here, \emph{competitive ability} determines the outcome of a pairwise
contest, whereas \emph{fitness} refers to demographic outcomes;
differences in competitive ability can contribute to average fitness
differences in the sense of \cite{Chesson2000}. Even under a common
constraint on mean competitive ability, distributions with the same mean can
perform differently against a given competitor. This allows us to
study the effects of individual variability without assigning ability
to a particular biological trait.  Related approaches include the open-site competition model of
Lichstein et al.~\cite{Lichstein2007Intraspecific} and Hart et al.'s
analysis of nonlinear averaging and stochastic demography
\cite{Hart2016IndividualVariation}.

The expected outcome of competition is only one part of the ecological problem. An advantage in individual encounters must be translated into changes in abundance, and random demographic events may cause a species to disappear despite such an advantage. Immigration introduces a further process by connecting local dynamics to the composition of a regional species pool. Recent work has extended the distribution-based game framework to adaptive dynamics, allowing strategies themselves to evolve \cite{KarlssonGerleeRowlett2025}. Here, we address a complementary question on an ecological time scale: when competitive-ability distributions are held fixed, how do their differences affect a species' stochastic abundance dynamics, probability of fixation, and abundance distribution in the presence of immigration?

We develop a finite-population Markov model in which a focal species competes with the aggregate of all other species in a local community of fixed size. Encounters between individuals of the focal species and this local aggregate determine competitive outcomes. Immigration also changes local abundance, with the probability of immigration and the relative abundance within the metacommunity determining how often immigrants arrive and which species they belong to. From these update rules, we derive a stochastic differential equation that separates systematic abundance change from demographic fluctuations. The difference between win and loss probabilities determines directional competition, while their sum determines the contribution of local competition to stochastic abundance changes. The model recovers neutral local-community dynamics as a special case and allows us to identify which consequences of individual variability arise through each component.

We analyse the model both without immigration and with continued input from the metacommunity. For a closed community, we obtain fixation probabilities in the diffusion approximation and examine how competitive advantage interacts with community size and demographic stochasticity. With immigration, we study probability-density dynamics and stationary abundance distributions under zero-flux boundary conditions. Constructed examples demonstrate that populations with identical mean competitive ability and variance can nevertheless have different competitive outcomes and demographic prospects against the same competitor. Our aim is to determine how information contained in the shape of an individual-level distribution propagates into population dynamics, and when replacing that distribution by its mean conceals ecologically consequential differences. 

\section{Methods}\label{s:methods}

\begin{figure}[!htbp]
\centering
\includegraphics[width=\linewidth]{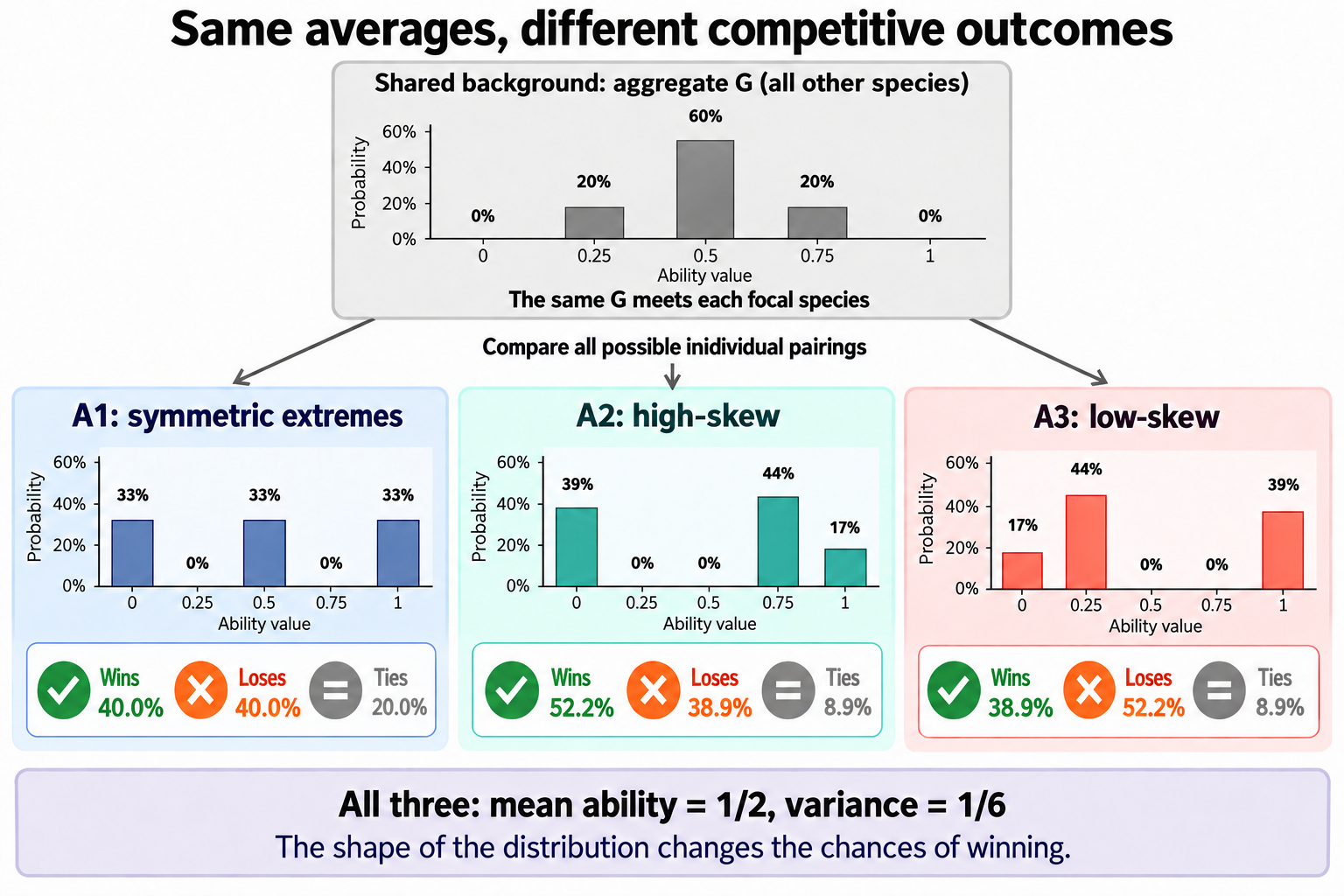}
\caption{Individual variability produces different competitive outcomes.
The three focal species $A_1$, $A_2$, and $A_3$ have the same mean
competitive ability, $1/2$, and variance, $1/6$, but different
distributions. Each competes against the same aggregate $\cG$.
Win, loss, and tie probabilities are calculated by independently drawing one individual from the focal species and one from the shared background aggregate species, comparing their competitive ability values, and cumulatively assessing the results over all pairings.
Displayed percentages are rounded.}
\label{fig:individual-variability}
\end{figure}

We extend Hubbell's fixed-size replacement framework \cite{Hubbell2001} by letting individual competitive-ability distributions determine encounter probabilities. Our contribution is to translate these outcomes into stochastic fixation and abundance dynamics. Shape adds demographic information when it changes win, loss, or tie probabilities; distributions yielding the same probabilities generate the same abundance process. A schematic overview of our model is shown in Figure \ref{fig:individual-variability}.  

Our model deliberately does not specify which particular trait or phenotype varies among individuals. Instead, individual variability is represented abstractly, so that the same framework can capture different biological mechanisms in different contexts. This flexibility is important because intraspecific variability can influence species performance through diverse pathways, including effects on growth, survival, competition, and environmental responses. Thus, the same mathematical framework can be applied to different biological systems by interpreting the variable individual characteristic according to the specific context.

\subsection{From variable individuals to competing species}\label{ss:markovgame}
Classical neutral theory assumes ecological equivalence among individuals with respect to their per-capita demographic rates \cite{Hubbell2001}. Here, we relax this assumption by allowing individuals to differ in their competitive characteristics. These characteristics could represent an aggregate over several phenotypes or biological mechanisms, or they could represent a single specific trait. Each individual is assigned a \emph{competitive ability}, representing an abstract measure of its performance in pairwise competition. Competitive ability may reflect differences in traits affecting processes such as resource acquisition, survival, or reproductive success, without requiring that these mechanisms be modeled separately.

At each encounter, we independently sample a competitive ability for each participant from the distribution associated with its species. This encounter-level assumption represents variation in individual performance over time and across ecological contexts, motivated by observed fluctuations in single-cell growth and changes in microbial interactions \cite{Kiviet2014,Daniels2023}. Successive draws are assumed independent for tractability. 

We consider a species $A$ competing with all other species, which we collectively represent by a single aggregate competitor $\cG$. We represent all non-focal individuals, including immigrants, by a fixed competitive-ability distribution for \(\mathcal G\). This is exact if the non-focal species share that distribution; their combined abundance can still change, although changes in their species composition are not tracked. 
We emphasize that $\cG$ consists of \emph{all other species in the local community}, and that the number of competing species is unrestricted. 

There are different ways one could compute the expected outcome of competition between $A$ and $\cG$. Here we present our method, developed in \cite{MendenDeuerRowlett2014,MendenDeuerRowlett2019,MendenDeuerRowlett2021,RowlettKarlssonNursultanov2022}. In the discrete formulation, the possible competitive abilities are $\frac jM$ for $j=0,\ldots,M$, for a fixed integer $M>0$. The value $M$ determines the resolution of the trait space and may be freely selected according to the context.
Therefore, the species is not characterized by a single competitive ability, but by the distribution of competitive abilities among its individuals. We refer to this distribution as the species' \emph{strategy}. An example of three different focal species in competition with the same aggregate competitor is shown in Figure \ref{fig:individual-variability}.

Let $F_A$ and $F_\cG$ denote the competitive abilities of independently sampled individuals from $A$ and $\cG$, respectively. In the discrete formulation, their distributions are specified by
\[
a_j=\Pr\left(F_A=\frac jM\right),\quad \sum_{j=0}^{M}a_j=1, \qquad
g_j=\Pr\left(F_\cG=\frac jM\right),\qquad \sum_{j=0}^{M}g_j=1.
\]
The mean competitive abilities are respectively 
\beq
\operatorname{MCA}(A)=\sum_{j=0}^{M}\frac jM a_j, \qquad 
\operatorname{MCA}(\cG)=\sum_{j=0}^{M}\frac jM g_j.
\label{eq:mca} 
\eeq 
Both are subject to the common constraint
\[\operatorname{MCA}(A)\leq\frac12 \quad \operatorname{MCA}(\cG) \leq \frac 1 2.\]
This constraint prevents competitive advantage from being obtained simply by shifting all individuals toward arbitrarily high competitive abilities, as discussed in \cite{RowlettKarlssonNursultanov2022}, and instead makes the shape of the intraspecific distribution a key factor in determining competitive success.

When an individual from species $A$ encounters an individual from a different species in the local community, the individual with the greater competitive ability wins the encounter, whereas equal competitive abilities result in a tie. These individual-level encounters determine the expected competitive relationship between the species and, ultimately, their stochastic abundance dynamics in the local community. If the individual drawn from species $A$ has competitive ability $j/M$, then it defeats every individual having competitive ability strictly smaller than $j/M$. Hence, conditioned on selecting an individual with competitive ability $j/M$, the probability of winning in competition with the aggregate local community $\cG$ is
\[\sum_{i<j}g_i.\]
Since the probability of selecting such an individual is $a_j$, the law of total probability gives
\begin{equation}
W_{A\cG}=\sum_{j=0}^{M}a_j\sum_{i<j}g_i=\Pr(F_A>F_\cG).
\label{eq:winsdiscrete}
\end{equation}
Species $A$ loses whenever the selected individual from $\cG$ has strictly larger competitive ability. Thus,
\begin{equation}
L_{A\cG}=\sum_{j=0}^{M}a_j\sum_{i>j}g_i=\Pr(F_A<F_\cG).
\label{eq:lossesdiscrete}
\end{equation}
If both selected individuals possess identical competitive ability, neither species gains a competitive advantage. The probability of this event is therefore
\beq T_{A\cG}=\sum_{j=0}^{M}a_jg_j=\Pr(F_A=F_\cG).\label{eq:tiesdiscrete} \eeq 
Since exactly one of the three mutually exclusive events must occur,
\[W_{A\cG}+L_{A\cG}+T_{A\cG}=1.\]
We could also define these values using the continuous game-theoretic model, in which the strategies are represented by bounded probability densities also denoted $F_A$ and $F_\cG$ on $[0,1]$.  In this case, competitive abilities can be any real number between $0$ and $1$, and the probabilities are
\begin{align}
W_{A\cG}&=\int_0^1 F_A(x)\left(\int_0^x F_\cG(y)\,dy\right)\,dx=\Pr(F_A>F_\cG), \quad \int_0 ^1 F_A (x) dx = 1, \label{eq:wincont} \\
L_{A\cG}&=\int_0^1 F_A(x)\left(\int_x^1 F_\cG(y)\,dy\right)\,dx=\Pr(F_A<F_\cG), \quad \int_0 ^1 F_\cG(x)dx = 1.\label{eq:lose_cont}
\end{align}
The MCA in this case is defined to be 
\beq \mca(A) = \int_0 ^1 x F_A (x) dx, \quad \mca(\cG) = \int_0 ^1 x F_\cG (x) dx, \label{eq:mca_cont} \eeq 
and is subject to the same constrained maximum value $\frac 1 2$. 
 
Either formulation of the game determines the parameters that appear in the Markov chain and the associated stochastic differential equation:
\begin{equation}
D_{A\cG}=W_{A\cG}-L_{A\cG},\qquad S_{A\cG}=W_{A\cG}+L_{A\cG}.\label{eq:wins_losses}
\end{equation}

In the discrete model, equal abilities represent genuine ties and produce no replacement. Numerical bins used to approximate continuous abilities should not be interpreted as biological ties.
To determine how individual variability affects population dynamics, we now connect the competitive outcomes defined above to changes in local species abundance. The probabilities $W_{A\cG}$ and $L_{A\cG}$ describe the outcome of an encounter, but their demographic consequences also depend on how frequently individuals of the focal species encounter competitors. We therefore construct a finite-population Markov chain in which competitive encounters lead to the replacement of one individual by another. This connects the distributions of individual competitive abilities to both the expected change in species abundance and the fluctuations arising from demographic stochasticity. We then derive a diffusion approximation and incorporate immigration from a regional metacommunity to obtain the full stochastic differential equation. Throughout this construction, we hold the competitive-ability distributions fixed while allowing abundances to change. For the aggregate competitor, this amounts to assuming that its effective competitive-ability distribution remains unchanged over the time scale considered.

\begin{table}[htbp]
\centering
\caption{Notation and abbreviations in the derivation of the stochastic differential equation.} 

\renewcommand{\arraystretch}{1.15}
\begin{tabularx}{\linewidth}{@{}l>{\raggedright\arraybackslash}X@{}}
\toprule
$A$ & Focal species. \\  
$\cG$ & Aggregate of all other species in the local community.\\
$M$ & Resolution of the discrete competitive ability space.\\
$F_A$ &  Competitive ability distribution or \em strategy \em of the focal species. \\ 
$F_\cG$ & Competitive ability distribution or \em strategy \em of the aggregate of all other species in the local community. \\ 
$a_j,g_j$ & Probabilities of competitive ability $j/M$ in $A$ and $\cG$, respectively, in the discrete competitive ability space. \\
$\operatorname{MCA}$ & Mean competitive ability defined in \eqref{eq:mca}, \eqref{eq:mca_cont}, for the discrete and continuous game formulations, respectively.\\
$W_{A\cG},L_{A\cG},T_{A\cG}$ & Probabilities that $A$ wins, loses, or ties in an encounter with $\cG$, defined according to the discrete or continuous games, respectively in \eqref{eq:winsdiscrete}, \eqref{eq:lossesdiscrete}, \eqref{eq:tiesdiscrete}, \eqref{eq:wincont}, \eqref{eq:lose_cont}.
\\
$D_{A\cG},S_{A\cG}$ & Difference and sum of win and loss probabilities given in equation~\eqref{eq:wins_losses}.\\
$J$ & Fixed number of individuals in the local community.\\
$A(k)$ & Abundance of $A$ after the $k$-th ecological update.\\
$h_n$ & Probability of sampling an $A$--$\cG$ pair given in equation~\eqref{eq:heterospecific_probability}.\\
$c$ & Probability that a sampled $A$--$\cG$ encounter results in a competition.\\
$p_n^+,p_n^-,p_n^0$ & Closed-community transition probabilities given respectively in \eqref{eq:markov_pplus}, \eqref{eq:markov_pminus}, and \eqref{eq:markov_pzero}.\\
$X_k,\,x=n/J$ & Discrete relative abundance and its current value.\\
$h_J(x)$ & Encounter probability as a function of relative abundance; equation~\eqref{eq:hJ}.\\
$\mu(x)$ & Drift coefficient with and without immigration, respectively, \eqref{eq:full_drift}, \eqref{eq:driftcoef1}. \\ 

$\sigma^2(x) $ & Squared diffusion coefficient with and without immigration, respectively,\eqref{eq:full_diffusion}, \eqref{eq:diffcoeff1}.\\
$X_t$ & Continuous diffusion approximation to the relative abundance of $A$.\\
$W_t$ & Standard Wiener process.\\
$m$ & Probability of an immigration event at each update.\\
$\pi_A$ & Relative abundance of $A$ in the metacommunity.\\
$q_n^+,q_n^-,q_n^0$ & Transition probabilities including immigration, given respectively in \eqref{eq:full_pplus}, \eqref{eq:full_pminus}, \eqref{eq:full_pzero}.\\
\bottomrule
\label{tabel_methods}
\end{tabularx}
\end{table}

\subsection{Finite-population Markov chain}\label{sec:markov-chain}
We consider a local ecological community containing a fixed number $J\geq2$ of individuals. This fixed-community-size 
assumption represents a saturated local community in which gains in the abundance of one species must be accompanied by corresponding losses elsewhere in the community \cite{Hubbell2001}. Each replacement event therefore preserves the total community size. 

Let
\[A(k)=n,\qquad \cG(k)=J-n,\]
denote the abundances of $A$ and its aggregate competitor $\cG$, respectively, after the $k$-th ecological update. Since
\[A(k)+\cG(k)=J,\]
it is sufficient to keep track of $A(k)$. The state space of the process is $ \{0,1,\ldots,J\}$.  
Thus, the state $n$ completely determines the abundance composition of the two-type local community.

We first consider a closed local community without immigration as in \cite{Hubbell2001} and subsequently include immigration from a larger metacommunity that contains the local community. 
At each local update, two individuals are sampled uniformly without replacement. Conditional on $A(k)=n$, the probability that one belongs to $A$ and the other to $\cG$ is
\begin{equation}
h_n =\frac nJ\frac{J-n}{J-1}+\frac{J-n}J\frac n{J-1}=\frac{2n(J-n)}{J(J-1)}.\label{eq:heterospecific_probability}
\end{equation}
The two terms correspond to sampling an $A$-individual followed by a $\cG$-individual, or a $\cG$-individual followed by an $A$-individual, respectively. The factor $h_n$ therefore measures how frequently individuals from the two types encounter one another under random sampling. In particular,
\[h_0=h_J=0.\]

In the classical neutral model individuals are assumed to be
well mixed: any two individuals are equally likely to interact. In natural communities, incomplete mixing, spatial aggregation, or habitat structure reduce the rate at which individuals of different types actually encounter one another, and it is well known that such spatial structure weakens interspecific competition
\cite{DurrettLevin1994, Tilman1994, BolkerPacala1997, Chesson2000}.  We account for this in a mean-field way by introducing an interaction parameter $c\in[0,1]$, defined as the probability that an $A$--$\mathcal G$ pair selected by the sampling rule actually interacts, relative to the well-mixed case. Thus $ch_n$ is the probability that a local update results in a competitive interaction between $A$ and $\mathcal G$; with probability $1-ch_n$ no competitive interaction occurs.  The case $c=1$ recovers complete mixing, while $c<1$ reflects reduced encounter rates, for instance when individuals compete only within patches and $c$ is the probability that a sampled pair shares a patch.

An increase in the abundance of $A$ occurs when an $A$--$\cG$ pair is selected,
and the $A$-individual wins the competition. Hence,
\begin{equation}
p_n^+=\Pr\!\left(A(k+1)=n+1\mid A(k)=n\right)=ch_nW_{A\cG}.\label{eq:markov_pplus}
\end{equation}
Similarly, the abundance of $A$ decreases when the $\cG$-individual wins, giving
\begin{equation}
p_n^-=\Pr\!\left(A(k+1)=n-1\mid A(k)=n\right)=ch_nL_{A\cG}.\label{eq:markov_pminus}
\end{equation}
All remaining events leave the abundance of $A$ unchanged. Therefore,
\begin{align}
p_n^0&=\Pr\!\left(A(k+1)=n\mid A(k)=n\right)\nonumber\\
&=1-p_n^+-p_n^-\nonumber\\
&= 1-ch_n S_{A\cG}, \quad S_{A\cG} := W_{A\cG} + L_{A \cG} .\label{eq:markov_pzero}
\end{align}

Consequently, the abundance process can be written, conditional on $A(k)  =n$, as 
\begin{equation}
\begin{gathered}
A(k+1)=\begin{cases}
n+1,&\text{with probability }p_n^+,\\[4pt]
n-1,&\text{with probability }p_n^-,\\[4pt]
n,&\text{with probability }p_n^0,
\end{cases}\\
\end{gathered}\label{eq:birth_death_chain}
\end{equation}
The resulting process is therefore a finite one-dimensional birth--death Markov chain.

\subsubsection{Directional competition and demographic variability}
The transition probabilities naturally separate two different consequences of the competitive-ability distributions. The difference between the upward and downward transition probabilities is
\begin{equation}
p_n^+-p_n^-=ch_nD_{A\cG},\label{eq:transition_difference}
\end{equation}
whereas their sum is
\begin{equation}
p_n^++p_n^-=ch_nS_{A\cG}.\label{eq:transition_sum}
\end{equation}
These two quantities have distinct ecological interpretations. The quantity $D_{A\cG}$ determines the directional bias of competition: if $D_{A\cG}>0$, competitive interactions favor species $A$ on average, whereas $D_{A\cG}<0$ favors $\cG$. By contrast, $S_{A\cG}$ determines how frequently competitive encounters are decisive and therefore controls the intensity of abundance fluctuations.

This distinction becomes especially clear by defining the one-step abundance increment
\[\Delta A(k)=A(k+1)-A(k).\]
Conditional on $A(k)=n$,
\[
\Delta A(k)=\begin{cases}
1,&\text{with probability }p_n^+,\\
-1,&\text{with probability }p_n^-,\\
0,&\text{with probability }p_n^0.
\end{cases}
\]
Its first conditional moment is
\begin{equation}
\mathbb E\left[\Delta A(k)\mid A(k)=n\right]=ch_nD_{A\cG},\label{eq:first_jump_moment_N}
\end{equation}
while its second conditional moment is
\begin{equation}
\mathbb E\left[(\Delta A(k))^2\mid A(k)=n\right]=ch_nS_{A\cG}.\label{eq:second_jump_moment_N}
\end{equation}
The first quantity measures the systematic expected change in abundance, whereas the second measures the magnitude of stochastic abundance changes. After rescaling abundance and time, these two quantities will generate the drift and diffusion coefficients of the continuous approximation derived below.

\subsection{Diffusion approximation}\label{sec:diffusion_approximation}
The finite Markov chain provides an exact description of the abundance dynamics under the update rules specified above. To obtain a continuous approximation, we now consider the relative abundance of species $A$,
\[X_k=\frac{A(k)}J.\]
If $A(k)=n$, we write
\[x=\frac nJ, \textrm{ so that }  X_k\in\left\{0,\frac1J,\frac2J,\ldots,1\right\}.\]
Because a single transition changes the abundance of species $A$ by at most one individual, the corresponding change in relative abundance is
\[\Delta X_k=X_{k+1}-X_k=\frac{\Delta A(k)}J.\]
Hence,
\[
\Delta X_k=\begin{cases}
\frac1J,&\text{with probability }p_n^+,\\[6pt]
-\frac1J,&\text{with probability }p_n^-,\\[6pt]
0,&\text{with probability }p_n^0.
\end{cases}
\]
The first conditional moment is therefore
\begin{equation}
\mathbb E\left[\Delta X_k\mid X_k=x\right] =\frac1J p_n^+-\frac1J p_n^- 
 =\frac1J\left(p_n^+-p_n^-\right) =\frac cJ h_nD_{A\cG}.\label{eq:first_jump_moment_X}
\end{equation}
Similarly, because
\[
(\Delta X_k)^2=\begin{cases}
\frac1{J^2},&\text{if the abundance changes},\\[4pt]
0,&\text{otherwise},
\end{cases}
\]
the second conditional moment is
\begin{equation}
\mathbb E\left[(\Delta X_k)^2\mid X_k=x\right] =\frac1{J^2}\left(p_n^++p_n^-\right) =\frac c{J^2}h_nS_{A\cG}.\label{eq:second_jump_moment_X}
\end{equation}
To express these quantities directly in terms of relative abundance, we substitute $n=Jx$ into the probability of an encounter between $A$ and $\cG$:

\begin{equation} h_n=\frac{2n(J-n)}{J(J-1)} =\frac{2(Jx)(J-Jx)}{J(J-1)}  =\frac{2J}{J-1}x(1-x) = h_J(x). \label{eq:hJ}
\end{equation}

Here $h_J(x)$ is a function of relative abundance, whereas $h_n$ as we defined it in \eqref{eq:heterospecific_probability} denotes the discrete encounter probability. The factor $x(1-x)$ has a direct ecological interpretation. Encounters between the two types are most frequent when both occur at substantial abundance and become increasingly rare as either approaches local extinction. In particular, $h_J(0)=h_J(1)=0.$

\subsubsection{Ecological time scale}

We take update attempts to occur at rate \(J\), so that one ecological
time unit contains \(J\) attempts on average. The waiting time between
attempts is random, with mean \(1/J\). The expected change in relative abundance per unit ecological time is then
\begin{align}
\frac{\mathbb E\left[\Delta X_k\mid X_k=x\right]}{\Delta t} =J\mathbb E\left[\Delta X_k\mid X_k=x\right]=ch_J(x)D_{A\cG}.
\end{align}
This defines the drift coefficient
\begin{equation}
\mu(x)=ch_J(x)D_{A\cG}.\label{eq:mu_closed}
\end{equation}
Similarly, the second conditional jump moment per unit ecological time is
\begin{align}
\frac{\mathbb E\left[(\Delta X_k)^2\mid X_k=x\right]}{\Delta t}
=J\mathbb E\left[(\Delta X_k)^2\mid X_k=x\right] =\frac cJ h_J(x)S_{A\cG}.
\end{align}
This gives the squared diffusion coefficient in the second-order diffusion approximation:
\begin{equation}
\sigma^2(x)=\frac cJ h_J(x)S_{A\cG}.\label{eq:sigma_closed}
\end{equation}
Using these first two conditional jump moments gives the diffusion approximation
\begin{equation}
dX_t=\mu(X_t)\,dt+\sigma(X_t)\,dW_t,\label{eq:general_SDE}
\end{equation}
where $W_t$ is a standard Wiener process. 
For a large local community, $J/(J-1)\approx1$, and the diffusion approximation takes the simpler form
\begin{equation}
dX_t\approx2cD_{A\cG}X_t(1-X_t)\,dt+\sqrt{\frac{2cS_{A\cG}}J X_t(1-X_t)}\,dW_t,\label{eq:closed_SDE_largeJ}
\end{equation}
so in this formulation the drift coefficient  
\beq \mu(x) = 2 c D_{A \cG} x (1-x), \label{eq:driftcoef1} 
\eeq 
and squared diffusion coefficient 
\beq \sigma^2(x) = \frac{2 c S_{A \cG}}{J} x (1-x). \label{eq:diffcoeff1} \eeq

\subsection{Immigration and the full stochastic differential equation}
\label{sec:immigration}

\begin{figure}[h]
    \centering
    \includegraphics[width=0.75\linewidth]{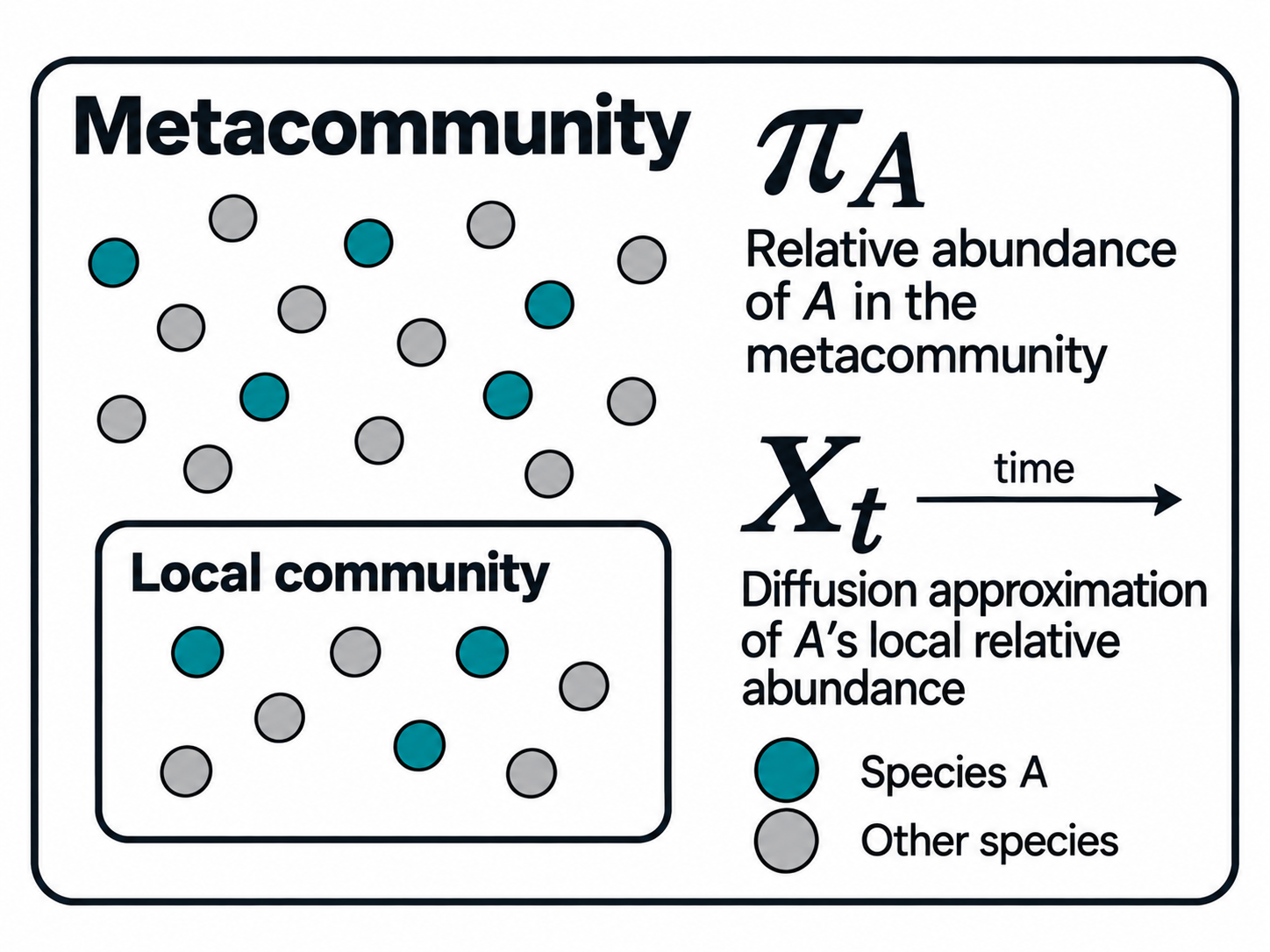}
    \caption{Schematic illustration of the local community contained within a larger metacommunity.  Teal circles represent individuals of the focal species \(A\);    gray circles represent individuals of other species.
    The parameter \(\pi_A\) is the relative abundance of \(A\)
    within the metacommunity, while \(X_t\) is the continuous
    diffusion approximation of its relative abundance in the
    local community at time \(t\). Circle counts are illustrative.}
    \label{fig:meta-local}
\end{figure}

The closed-community model describes changes in abundance caused by local competitive interactions. In the absence of immigration, the states $X_t=0$ and $X_t=1$ are absorbing. They correspond respectively to local extinction and fixation of species $A$.  We now extend this model to account for immigration from a regional metacommunity. Immigration provides an additional source of individuals and can therefore restore a species that has disappeared locally, provided that it remains present in the metacommunity. This is depicted in Figure \ref{fig:meta-local}.

We retain the assumption that the local community contains a fixed number $J$ of individuals. At each update, an immigration event occurs with probability $m\in[0,1]$. During such an event, one individual is selected uniformly from the local community and replaced by an immigrant. With the remaining probability $1-m$, the update follows the local competition mechanism described above. Thus, immigration and local competition are alternative update mechanisms, both of which preserve the total community size.

Let $\pi_A\in[0,1]$ denote the relative abundance of species $A$ in the metacommunity. We assume that immigrants are sampled according to metacommunity relative abundances, so an immigrant belongs to $A$ with probability $\pi_A$ and to another species with probability $1-\pi_A$. Here $m$ and $\pi_A$ are treated as constant parameters, representing an immigration regime and a metacommunity composition that remain approximately unchanged over the time scale considered as in \cite{Hubbell2001}.

\subsubsection{Transition probabilities with immigration}
Conditional on $A(k)=n$, an immigration event increases the abundance of $A$ when the individual being replaced belongs to $\cG$ and the immigrant belongs to $A$. The probability of this event is
\[m\frac{J-n}{J}\pi_A.\]
Conversely, the abundance of $A$ decreases when an $A$-individual is replaced by an immigrant belonging to another species. This occurs with probability
\[m\frac nJ(1-\pi_A).\]
If the immigrant and the replaced individual belong to the same type, the abundance of $A$ is unchanged.

To distinguish the full model from the closed-community model, write $q_n^+$, $q_n^-$, and $q_n^0$ for the transition probabilities including immigration. Combining the two update mechanisms gives
\begin{align}
q_n^+&=(1-m)p_n^++m\pi_A\frac{J-n}{J}\nonumber\\
&=(1-m)ch_nW_{A\cG}+m\pi_A\left(1-\frac nJ\right),\label{eq:full_pplus}\\
q_n^-&=(1-m)p_n^-+m(1-\pi_A)\frac nJ\nonumber\\
&=(1-m)ch_nL_{A\cG}+m(1-\pi_A)\frac nJ,\label{eq:full_pminus}\\
q_n^0&=1-q_n^+-q_n^-.\label{eq:full_pzero}
\end{align}
These probabilities define a birth--death Markov chain with self-loops on the same state space $\{0,1,\ldots,J\}$.

Writing $x=n/J$ and using $h_n=h_J(x)$, the difference between the upward and downward transition probabilities is
\begin{align}
q_n^+-q_n^-&=(1-m)ch_J(x)(W_{A\cG}-L_{A\cG})
+m\bigl[\pi_A(1-x)-(1-\pi_A)x\bigr]\nonumber\\
&=(1-m)ch_J(x)D_{A\cG}+m(\pi_A-x).\label{eq:full_transition_difference}
\end{align}
Their sum is
\begin{align}
q_n^++q_n^-&=(1-m)ch_J(x)(W_{A\cG}+L_{A\cG})
+m\bigl[\pi_A(1-x)+(1-\pi_A)x\bigr]\nonumber\\
&=(1-m)ch_J(x)S_{A\cG}
+m\bigl[\pi_A(1-x)+(1-\pi_A)x\bigr].\label{eq:full_transition_sum}
\end{align}
The difference determines the expected direction of abundance change. Its immigration contribution, $m(\pi_A-x)$, is positive when $A$ is less common locally than in the metacommunity and negative when it is more common locally. Immigration therefore tends to move local relative abundance toward the metacommunity value. The sum measures the probability that an update changes the abundance of $A$ and includes changes caused by both competition and immigration.

\begin{figure}[p]
\centering
\includegraphics[
  width=\linewidth,
  height=0.85\textheight,
  keepaspectratio
]{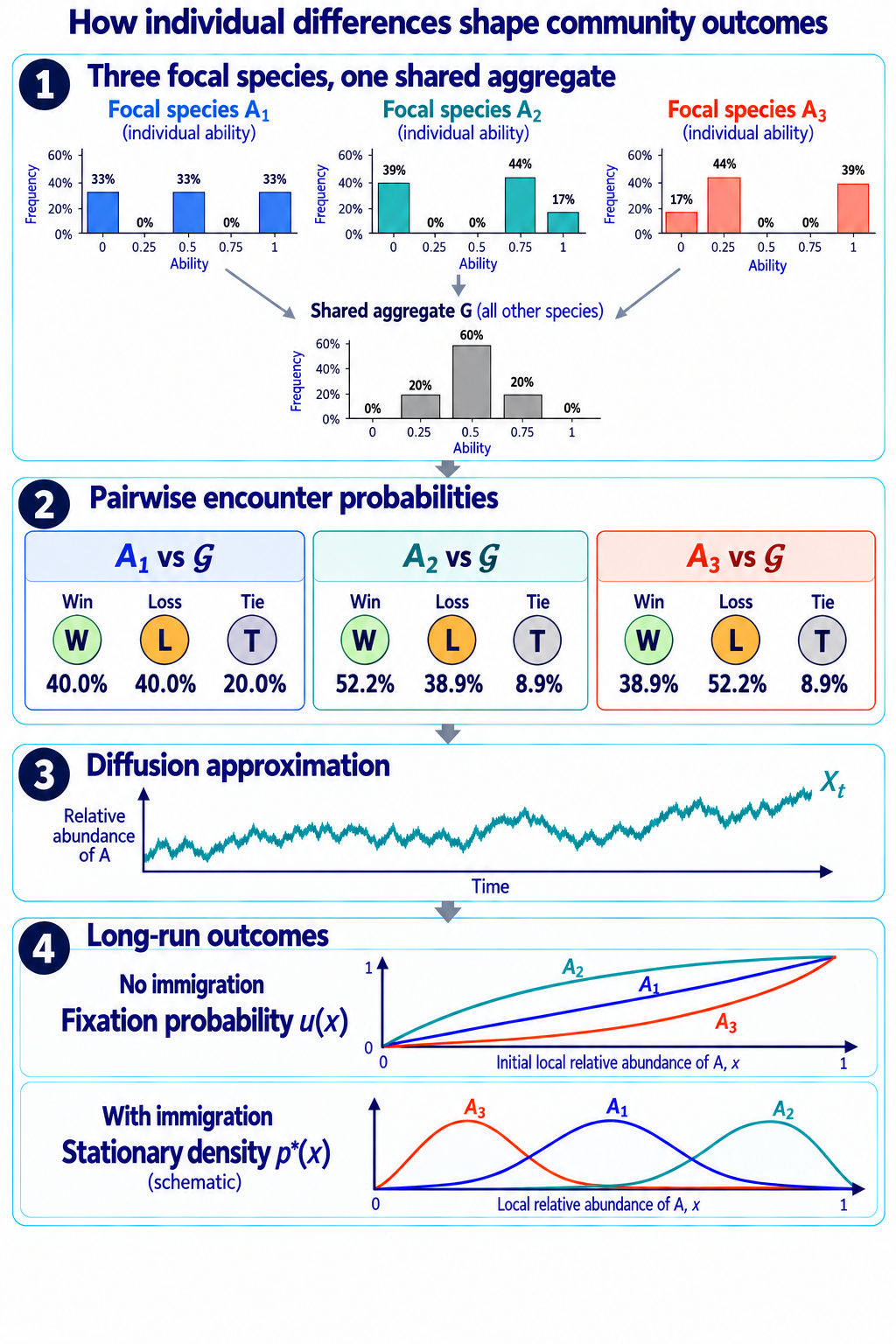}
\caption{(1) The three focal species $A_1,A_2,A_3$ have different
competitive-ability distributions, each with mean $1/2$ and
variance $1/6$. Each competes with the same aggregate
$\mathcal G$ of all other species.
(2) Pairwise encounters determine the probabilities of a win,
loss, or tie, denoted by $W$, $L$, and $T$. These enter the transition probabilities
of the finite-community model. 
(3) The model yields a diffusion approximation $X_t$ for the
focal species' local relative abundance.
(4) Without immigration, the corresponding fixation probabilities
are neutral for $A_1$, above neutral for $A_2$, and below neutral
for $A_3$. With immigration, the curves illustrate possible
stationary abundance densities; their shapes are schematic
rather than numerical solutions for a specified parameter set.}
\label{fig:individual-variability-schematic}
\end{figure}

\subsubsection{Diffusion coefficients}
As before, one update changes relative abundance by $1/J$, $-1/J$, or zero. Hence,
\begin{align}
\mathbb E\left[\Delta X_k\mid X_k=x\right]&=\frac1J(q_n^+-q_n^-),\label{eq:full_first_moment}\\
\mathbb E\left[(\Delta X_k)^2\mid X_k=x\right]&=\frac1{J^2}(q_n^++q_n^-).\label{eq:full_second_moment}
\end{align}
Using the same ecological time scale, $\Delta t=1/J$, the first conditional jump moment per unit time gives the drift coefficient
\begin{align}
\mu(x)&=\frac{\mathbb E[\Delta X_k\mid X_k=x]}{\Delta t}\nonumber\\
&=(1-m)ch_J(x)D_{A\cG}+m(\pi_A-x).\label{eq:full_drift}
\end{align}
The second conditional jump moment per unit time gives the squared diffusion coefficient in the second-order diffusion approximation:
\begin{align}
\sigma^2(x)&=\frac{\mathbb E[(\Delta X_k)^2\mid X_k=x]}{\Delta t}\nonumber\\
&=\frac1J\Bigl[(1-m)ch_J(x)S_{A\cG}
+m\bigl(\pi_A(1-x)+(1-\pi_A)x\bigr)\Bigr].\label{eq:full_diffusion}
\end{align}
The drift combines the directional effect of local competition with the tendency of immigration to move local composition toward that of the metacommunity. The squared diffusion coefficient combines demographic fluctuations associated with decisive competitive encounters and random immigrant replacements. Individual variability enters these coefficients through $D_{A\cG}$ and $S_{A\cG}$, thereby connecting the distribution of competitive abilities to the systematic and stochastic components of population dynamics.  Consequently, the full diffusion approximation is
\begin{equation}
\begin{aligned}
dX_t={}&\left[(1-m)ch_J(X_t)D_{A\cG}+m(\pi_A-X_t)\right]dt\\
&+\sqrt{\frac1J\Bigl[(1-m)ch_J(X_t)S_{A\cG}
+m\bigl(\pi_A(1-X_t)+(1-\pi_A)X_t\bigr)\Bigr]}\,dW_t.
\end{aligned}\label{eq:mastersde}
\end{equation}
The boundary conditions will be introduced in the following section when we formulate and study the associated Kolmogorov equations.

\section{Results}\label{s:results}

To study the stochastic dynamics and the role of individual variability on these dynamics, we solve the Kolmogorov equations associated with the SDE. To describe these equations, we note that the infinitesimal generator $\mathcal L$ of \eqref{eq:mastersde} acts on a test function $f$ via
\begin{equation}
\mathcal Lf(x)=\mu(x)f'(x)+\frac12\sigma^2(x)f''(x),\label{eq:fancyL}
\end{equation}
with coefficients
\begin{align}
\mu(x)&=(1-m)ch_J(x)D_{A\cG}+m(\pi_A-x),\label{eq:mu}\\
\sigma^2(x)&=\frac1J\Bigl[(1-m)ch_J(x)S_{A\cG}
+m\bigl(\pi_A(1-x)+(1-\pi_A)x\bigr)\Bigr].\label{eq:sig}
\end{align}
Here $x$ is the \emph{relative abundance} of type $A$, so the intended state space is the bounded interval $x\in[0,1]$. 
If $\rho(t,x)$ denotes the probability density of $X_t$, then in the interior of the state space $\rho$ satisfies
\begin{equation}\label{eq:fokker-planck}
\frac{\partial\rho}{\partial t}
=-\frac{\partial}{\partial x}\left(\mu(x)\rho(t,x)\right)
+\frac12\frac{\partial^2}{\partial x^2}\left(\sigma^2(x)\rho(t,x)\right)
=\mathcal L^{*}(\rho(t,x)).
\end{equation}
Equation~\eqref{eq:fokker-planck} is called the \emph{forward Kolmogorov equation}, or equivalently the \emph{Fokker--Planck equation} and evolves the probability distribution of the process.   The backward equation, \(\partial_t u=\mathcal L u\), evolves expected observables conditional on the initial state, and we solve both.

\begin{table}[htbp]
\centering
\caption{Notation and abbreviations in the solutions of the stochastic differential equation.} 

\renewcommand{\arraystretch}{1.15}
\begin{tabularx}{\linewidth}{@{}l>{\raggedright\arraybackslash}X@{}}
\toprule
$\cL$ & the infinitesimal generator of \eqref{eq:mastersde} defined in \eqref{eq:fancyL}. \\ 
$\mu(x)$ & Drift coefficient in \eqref{eq:mu}\\
$\sigma^2(x)$ & Squared diffusion coefficient in \eqref{eq:sig}.\\
$\cL^*$ & the adjoint of $\cL$ defined in \eqref{eq:fokker-planck}.\\ 
$\rho(x,t)$ & The probability density of $X_t$ which solves \eqref{eq:fokker-planck}. \\ 
$\wp$ & Dimensionless number defined in \eqref{eq:wp}. \\ 
$u(x)$ & Fixation probability in the absence of immigration which solves \eqref{eq:ode_trump} and is explicitly given in  \eqref{eq:soln_nodrift}. \\ 
$\mathcal J(t,x)$ & Probability current defined in \eqref{eq:prob_current}. \\ 
$\pstar(x)$ & Stationary density, satisfying \eqref{eq:pstar_eq1}, \eqref{eq:pstarbc1}, and \eqref{eq:pstarbc2}. \\
$r_{1,2}$ & Solutions to \eqref{eq:root-equation}, given in \eqref{eq:roots}, used to define the stationary density $\pstar$.  \\ 
$Z$ & Normalization constant in the stationary density to guarantee $\int_0 ^1 \pstar(x) dx = 1$. \\ 
$\lambda_j$, $j =0, 1, \ldots$, & Eigenvalues of the Sturm-Liouville problem for $\cL$ formulated in \eqref{eq:cL_pstar} and \eqref{eq:cL_slp}. \\ 

\bottomrule
\label{tabel_results}
\end{tabularx}
\end{table}

\subsection{Fixation probabilities in the absence of immigration}\label{s:trump}
Without immigration, $m=0$, and we have the simplifications
\[\mu(x)=ch_J(x)D_{A\cG},\quad\sigma^2(x)=\frac cJ h_J(x)S_{A\cG}.\]
Therefore, the infinitesimal generator is
\[\mathcal Lf(x)=ch_J(x)D_{A\cG}f'(x)+\frac c{2J}h_J(x)S_{A\cG}f''(x).\]
For $c>0$ and $S_{A\cG}>0$, by the Markov property, the probability of eventual fixation must be unchanged in expectation after an arbitrarily short initial time interval.  Consequently, the fixation probability satisfies the backward equation subject to the boundary conditions
\[\mathcal Lu(x)=0,\quad0<x<1,\qquad u(0)=0,\quad u(1)=1.\]

Hence, we have the boundary value problem
\beq ch_J(x)D_{A\cG}u'(x)+\frac c{2J}h_J(x)S_{A\cG}u''(x)=0,\quad u(0)=0,\quad u(1)=1. \label{eq:ode_trump} \eeq 
We define the dimensionless number
\begin{equation}\label{eq:wp}
\begin{gathered}
\wp=\frac{2JD_{A\cG}}{S_{A\cG}},\\
\wp>0\iff D_{A\cG}>0,\quad\wp<0\iff D_{A\cG}<0,\quad\wp=0\iff D_{A\cG}=0.
\end{gathered}
\end{equation}
This quantity compares directional game-theoretic drift with demographic stochasticity. Large positive values of $\wp$ correspond to a strong advantage for $A$, while large negative values correspond to a strong disadvantage. 

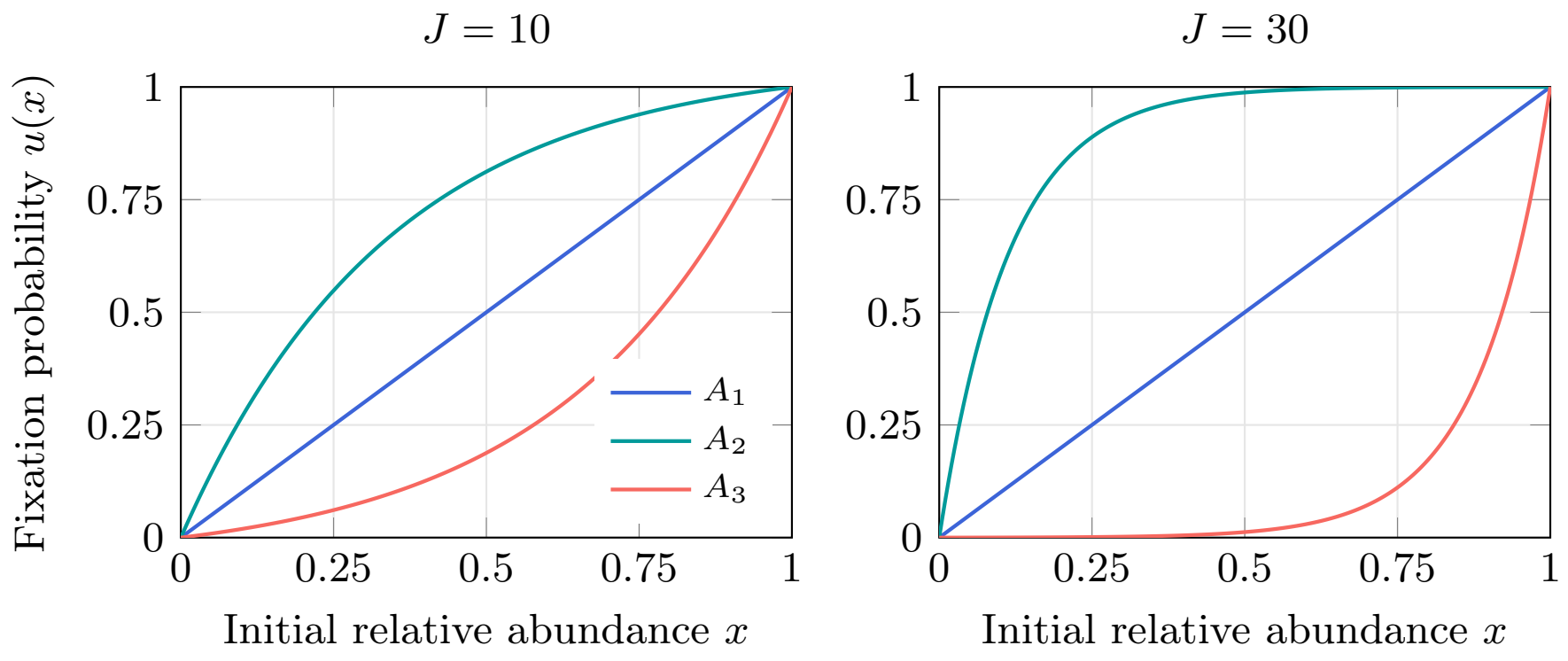
\begin{figure}[!htbp]
\centering
\begin{tikzpicture}
\begin{groupplot}[
  group style={group size=2 by 1, horizontal sep=1.1cm},
  width=0.41\linewidth,
  height=0.33\linewidth,
  xmin=0, xmax=1,
  ymin=0, ymax=1,
  xtick={0,0.25,0.5,0.75,1},
  ytick={0,0.25,0.5,0.75,1},
  xlabel={Initial relative abundance $x$},
  grid=major,
  grid style={gray!20},
  tick label style={font=\small},
  label style={font=\small},
  title style={font=\small}
]
\nextgroupplot[
  title={$J=10$},
  ylabel={Fixation probability $u(x)$},
  legend pos=south east,
  legend style={font=\scriptsize, draw=none, fill=white}
]
\addplot[Aone, thick, domain=0:1, samples=201] {x};
\addlegendentry{$A_1$}
\addplot[Atwo, thick, domain=0:1, samples=201]
  {(1-exp(-120*x/41))/(1-exp(-120/41))};
\addlegendentry{$A_2$}
\addplot[Athree, thick, domain=0:1, samples=201]
  {(exp(120*x/41)-1)/(exp(120/41)-1)};
\addlegendentry{$A_3$}

\nextgroupplot[title={$J=30$}]
\addplot[Aone, thick, domain=0:1, samples=201] {x};
\addplot[Atwo, thick, domain=0:1, samples=201]
  {(1-exp(-360*x/41))/(1-exp(-360/41))};
\addplot[Athree, thick, domain=0:1, samples=201]
  {(exp(360*x/41)-1)/(exp(360/41)-1)};
\end{groupplot}
\end{tikzpicture}
\caption{These are the fixation probabilities without immigration for the three
focal species in Figure~\ref{fig:individual-variability}.
Their competitive-advantage ratios are
$D_{A_1\cG}/S_{A_1\cG}=0$,
$D_{A_2\cG}/S_{A_2\cG}=6/41$, and
$D_{A_3\cG}/S_{A_3\cG}=-6/41$.
Increasing community size $J$ amplifies the departure from
neutrality for a fixed initial relative abundance $x$.}
\label{fig:fixation-probabilities}
\end{figure}

The solution is 
\beq
u(x)=
\begin{dcases}
\dfrac{1-\exp\!\left(-\dfrac{2JD_{A\cG}}{S_{A\cG}}x\right)}
      {1-\exp\!\left(-\dfrac{2JD_{A\cG}}{S_{A\cG}}\right)},
    & \wp \neq 0, \\[1ex]
x,  & \wp = 0.
\end{dcases}
\label{eq:soln_nodrift}
\eeq 
Thus,
\[\Pr\left(A\text{ fixates}\mid X_0=x\right)=u(x), \qquad \Pr\left(A\text{ becomes extinct}\mid X_0=x\right)=1-u(x).\]
If $D_{A\cG}>0$, then $\wp>0$ and
\[u(x)>x\qquad\text{for }0<x<1.\]
Thus, a positive value of $\wp$ increases the fixation probability of $A$ relative to the neutral value. Nevertheless, for finite $J$, $u(x)<1$ whenever $0<x<1$ and the advantage is finite. Therefore, a species $A$ consisting of variable individuals whose distribution has a competitive advantage makes survival more probable, although it does not guarantee survival. Fixation probabilities corresponding to the three focal species in Figure \ref{fig:individual-variability} are shown in Figure \ref{fig:fixation-probabilities}.

\subsubsection{Different competitive outcomes at identical mean competitive ability}
\label{sec:distribution_examples}
\begin{figure}[htbp]
\centering
\includegraphics[width=\linewidth]{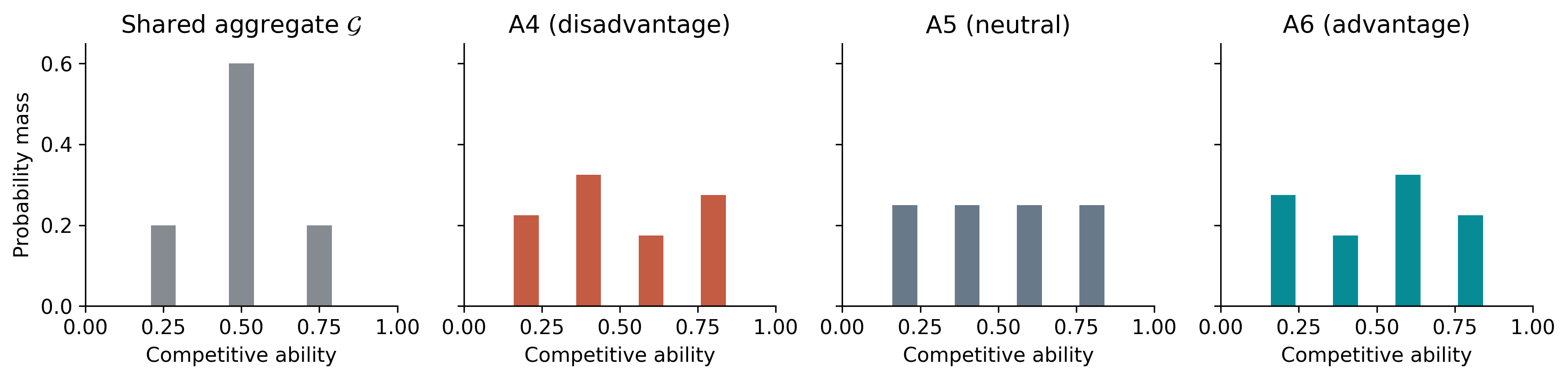}
\caption{Competitive-ability distributions for the shared aggregate
community $\cG$ and three focal species, $A_4$, $A_5$, and $A_6$.
The focal distributions all have mean $1/2$ and variance $1/20$,
but have different shapes. Against the same $\cG$, their respective
probabilities of winning an encounter are $0.48$, $0.50$, and $0.52$.  }
\label{fig:A4-A6-abilities}
\end{figure}

To further illustrate how individual variability affects competitive outcomes, we again consider three competitive-ability distributions for the focal species $A$, while keeping the aggregate competitor $\cG$ fixed.  In each of these distributions, the competitive ability takes values $0.2$, $0.4$, $0.6$, and $0.8$, with different probabilities as shown in 
(Figure~\ref{fig:A4-A6-abilities}). The individuals in the aggregate competitor $\cG$ take competitive ability values 0.25, 0.5, and 0.75, with probabilities shown in Figure \ref{fig:A4-A6-abilities}. 
The three distributions in Figure \ref{fig:A4-A6-abilities} have the same mean competitive ability, $0.5$, and the same variance, $0.05$. These values differ from those in Figure \ref{fig:individual-variability}, illustrating that our model applies to any distributions with any variance as long as the constraint on the mean competitive ability is satisfied. Consequently, differences in their competitive outcomes arise from aspects of distribution shape that are not captured by either the mean or the variance.

\begin{figure}[htbp]
\centering
\includegraphics[width=\linewidth]{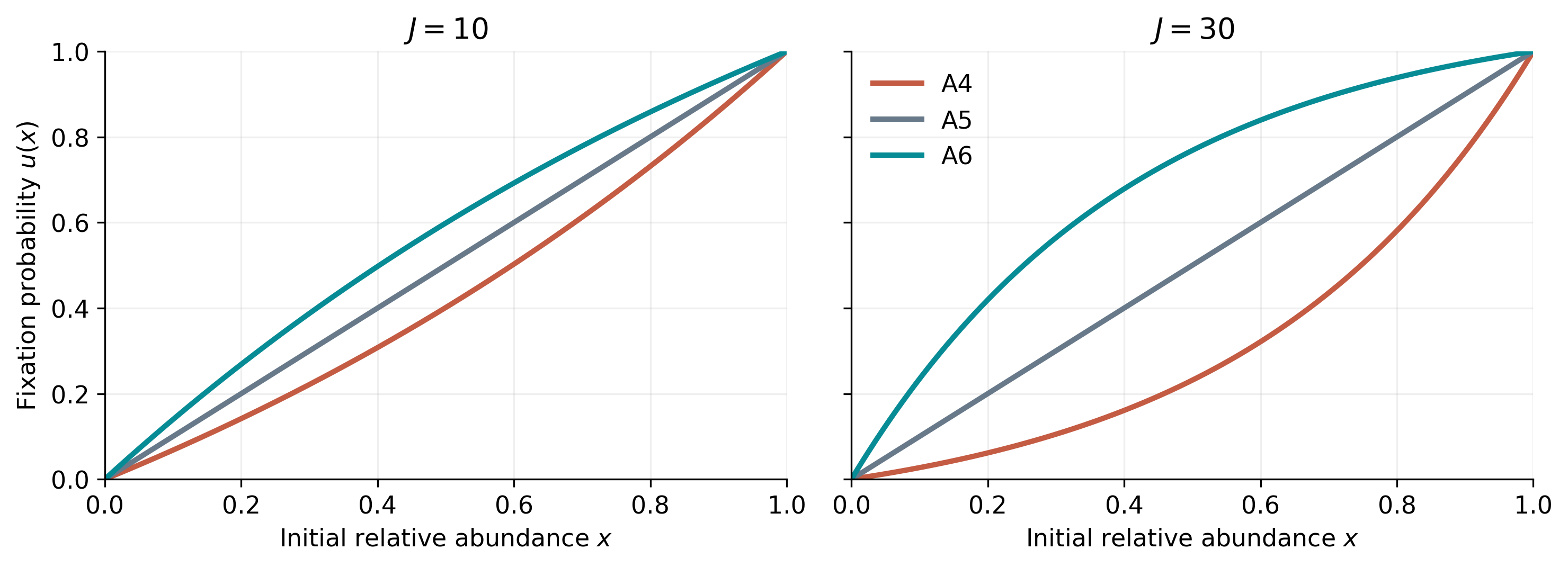}
\caption{Fixation probability $u(x)$ as a function of initial local
relative abundance $x$ in a closed community, for $J=10$ and $J=30$.
The three focal species have
$D_{A_i\cG}=-0.04,\,0,\,0.04$, respectively, and
$S_{A_i\cG}=1$. Thus $A_4$ is disadvantaged, $A_5$ is neutral,
and $A_6$ is advantaged against the shared aggregate $\cG$.}
\label{fig:A4-A6-fixation}
\end{figure}

Despite these shared characteristics, the probabilities of winning against $\cG$
are 
\[ D_{A_i\cG}=-0.04,\,0,\,0.04, \textrm{ respectively, for } i=4, 5, 6,\]  and
$S_{A_i\cG}=1$ for $i=4,5, 6$. Thus $A_4$ is disadvantaged, $A_5$ is neutral,
and $A_6$ is advantaged against the shared aggregate $\cG$.

Thus, distributions with identical mean competitive ability and variance
can produce a competitive disadvantage, neutrality, or a competitive
advantage against the same competitor
(Figure~\ref{fig:A4-A6-fixation}).

\subsection{Probability-density dynamics in the presence of immigration}\label{s:notrump}

With immigration, the boundary conditions are no longer absorbing.
The Fokker--Planck equation can be written as a continuity equation. Define the probability current
\begin{equation}
\mathcal J(t,x)=\mu(x)\rho(t,x)-\frac12\frac{\partial}{\partial x}\left(\sigma^2(x)\rho(t,x)\right). \label{eq:prob_current} 
\end{equation}
Then
\begin{equation}
\frac{\partial\rho}{\partial t}=-\frac{\partial\mathcal J}{\partial x}.
\end{equation}
Integrating over the state space gives
\begin{align}
\frac{d}{dt}\int_0^1\rho(t,x)\,dx
&=-\int_0^1\frac{\partial\mathcal J}{\partial x}\,dx\\
&=\mathcal J(t,0)-\mathcal J(t,1).
\end{align}
Therefore, we impose zero-flux boundary conditions
\begin{equation}\mathcal J(t,0)=\mathcal J(t,1)=0,\end{equation}
which guarantee that
\begin{equation}\frac{d}{dt}\int_0^1\rho(t,x)\,dx=0.\end{equation}
Consequently, the total probability is conserved. The operator $\cLstar$, with these boundary conditions, is the adjoint of the operator $\mathcal L$. In case $m=0$, there is no immigration, and fixation probabilities were obtained in \S\ref{s:trump}. Assuming $m>0$ and $0<\pi_A<1$, we have $\sigma^2(x)>0$ on the entire interval $[0,1]$, including the endpoints. Consequently, the operator $\cL$ is uniformly elliptic on this interval, and its boundary conditions are Neumann. 

\subsubsection{The stationary density} \label{s:stationary} 
For any initial condition, the solution of \eqref{eq:fokker-planck} converges exponentially to the stationary density, which satisfies 
\beq \cL^* \pstar(x) = 0, \qquad \pstar(x) \geq 0, \quad \int_0 ^1 \pstar(x) dx = 1, \label{eq:pstar_eq1} \eeq 
together with the zero-flux boundary conditions 
\beq 
m\pi_A \pstar(0)
-\frac{1}{2}
\left.\frac{d}{dx}\bigl(\sigma^2(x)\pstar(x)\bigr)\right|_{x=0}
&=0, \label{eq:pstarbc1} \\ 
-m(1-\pi_A)\pstar(1)
-\frac{1}{2}
\left.\frac{d}{dx}\bigl(\sigma^2(x)\pstar(x)\bigr)\right|_{x=1}
&=0. \label{eq:pstarbc2} 
\eeq 

We explicitly solve for the stationary density assuming 
\beq J>1, \qquad 0<m<1, \qquad cS_{A\mathcal G}>0, \qquad 0<\pi_A<1. \label{eq:imm_ass} \eeq 
This corresponds to a local community with more than one individual, with probability of immigration between $0$ and $1$, with competitive interactions occurring with probability greater than zero, and with the focal species $A$ present in, but not comprising all of, the metacommunity.  To solve this equation, we therefore find the roots of the quadratic equation: 
\begin{equation}
\frac{2J}{J-1}(1-m)cS_{A\cG}\,r(1-r)
+m\bigl[\pi_A+(1-2\pi_A)r\bigr]=0.
\label{eq:root-equation}
\end{equation}

The roots are 
\beq 
r_{1,2}&=&
\frac{
2J(1-m)cS_{A \cG}+(J-1)m(1-2\pi_A)}{4J(1-m)c S_{A \cG}}
\nn \\ 
& \mp & 
\frac{\sqrt{
\bigl[2J(1-m)cS_{A\cG}
      +(J-1)m(1-2\pi_A)\bigr]^2
+8J(J-1)m\pi_A(1-m)cS_{A\cG}
}
}{
4J(1-m)cS_{A\cG}
}.
\label{eq:roots}
\eeq 

\begin{figure}[htbp]
\centering
\includegraphics[width=0.85\textwidth]{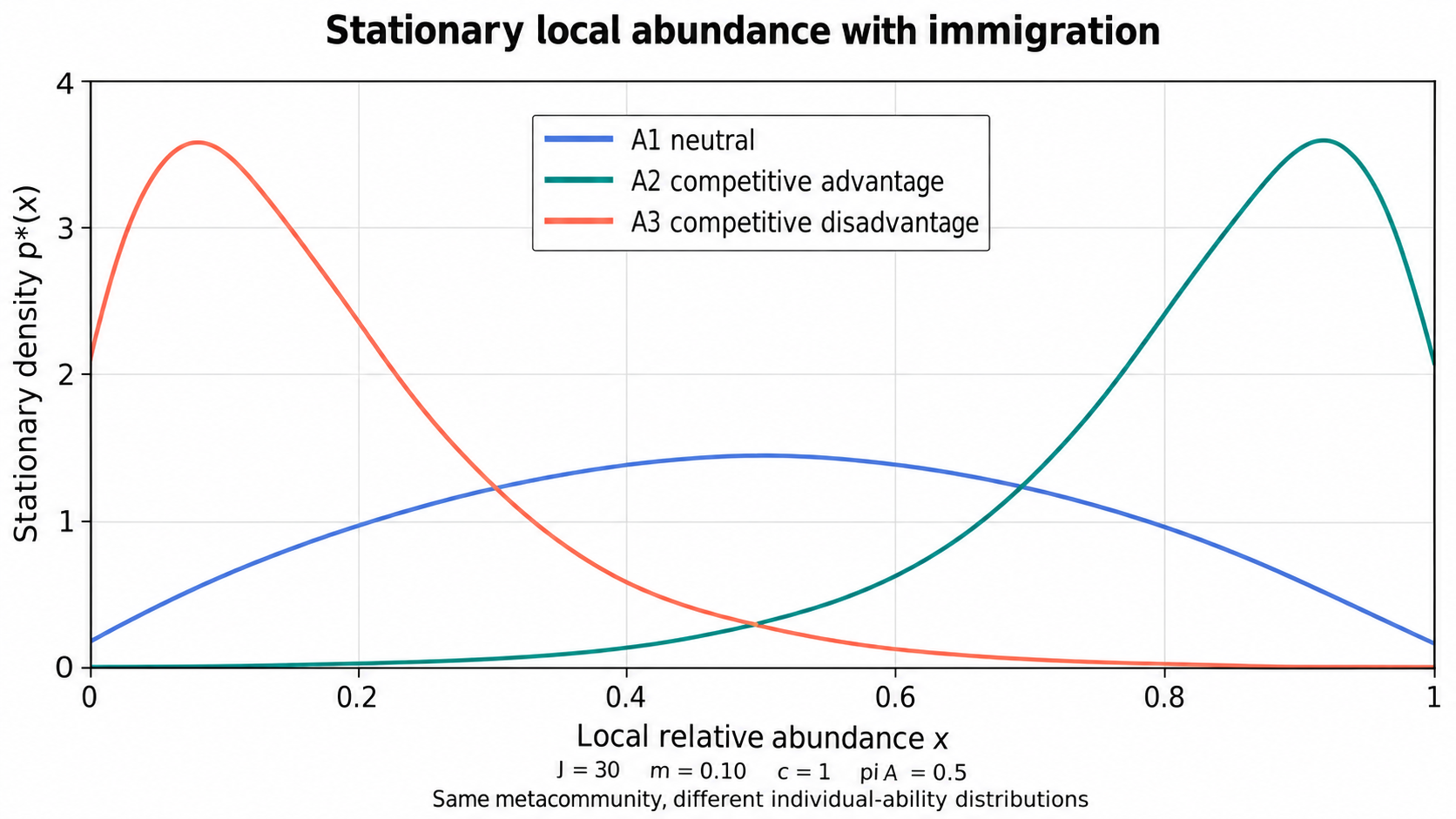}
\caption{Stationary probability densities of the local relative abundance of the focal species for the three competitive-ability distributions in Figure~\ref{fig:individual-variability}. All three focal species compete against the same aggregate community $\cG$. The parameters are $J=30$, $m=0.10$, $c=1$, and $\pi_A=1/2$. Although the three focal species have identical mean and variance in competitive ability, their different encounter outcomes produce different stationary abundance distributions.}\label{fig:stationary_density2}\end{figure}

\begin{figure}[htbp]
\centering
\includegraphics[width=0.85\linewidth]{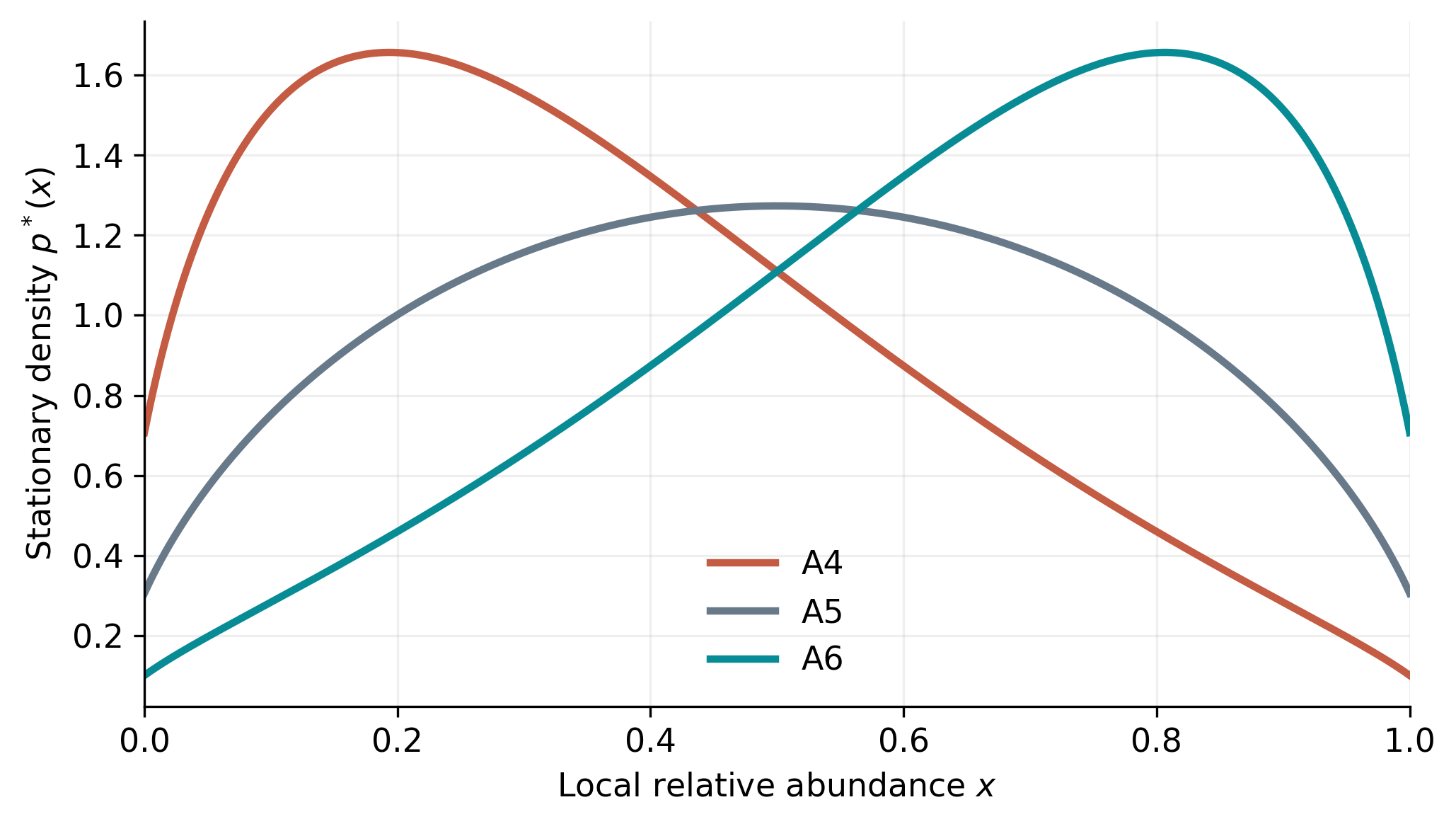}
\caption{Stationary probability densities of local relative abundance
for $A_4$, $A_5$, and $A_6$, from Figure \ref{fig:A4-A6-abilities} each competing against the same aggregate $\cG$. The parameters are $J=30$, $m=0.10$, $c=1$, and $\pi_A=1/2$. Although the focal species have identical means and variances in competitive ability, their different distribution
shapes lead to different stationary abundance distributions.}
\label{fig:A4-A6-stationary}
\end{figure}

With these values, we then have the stationary density given by 
\begin{equation}
\begin{aligned}
\pstar(x)=\frac{1}{Z}\,
&(x-r_1)^{
  \textstyle
  \frac{
    2J(1-m)cD_{A\cG}\,r_1(1-r_1)
    +(J-1)m(\pi_A-r_1)
  }{
    (1-m)cS_{A\cG}(r_2-r_1)
  }-1
}\\[2ex]
&\times(r_2-x)^{
  \textstyle
  -\frac{
    2J(1-m)cD_{A\cG}\,r_2(1-r_2)
    +(J-1)m(\pi_A-r_2)
  }{
    (1-m)cS_{A\cG}(r_2-r_1)
  }-1
}\\[2ex]
&\times
\exp\!\left(\frac{2JD_{A\cG}}{S_{A\cG}}x\right),
\qquad 0\leq x\leq1.
\end{aligned}
\label{eq:stationary-density}
\end{equation}
Here, the normalizing constant $Z$ is chosen so that the last condition on the integral of $\pstar$ in \eqref{eq:pstar_eq1} is satisfied.  Individual variability affects the local relative abundance in the stationary probability density, as can be seen in Figures \ref{fig:stationary_density2} and \ref{fig:A4-A6-stationary}.

\subsubsection{Limiting cases of the stationary density.}

If \(m=1\), the selection terms vanish. For \(0<\pi_A<1\) the normalized zero-flux density is
\begin{equation}
\pstar(x)=\frac{1}{Z}
\bigl[\pi_A+(1-2\pi_A)x\bigr]^{
\frac{4J\pi_A(1-\pi_A)}{(1-2\pi_A)^2}-1}
\exp\!\left(-\frac{2Jx}{1-2\pi_A}\right),
\qquad 0\leq x\leq1, \quad \pi_A \neq \frac 1 2,  \label{eq:m=1_not12} 
\end{equation}
\begin{equation}
\pstar(x)=\frac{1}{Z}
\exp\!\left[-2J\left(x-\frac12\right)^2\right],
\qquad 0\leq x\leq1, \quad \pi_A = \frac 1 2. \label{eq:m=1_12}
\end{equation}
In each expression, \(Z\) normalizes the density on \([0,1]\), so that it integrates to one. If $m=0$, then the situation reduces to the no-immigration case of \S \ref{s:trump}.  For $m > 0$ with $\pi_A = 0$, there is no immigration from the focal species, but immigration from other species still occurs.  The corresponding endpoint $x=1$ is not absorbing while $x=0$ is absorbing. If \(m>0\) and \(\pi_A=1\), the corresponding endpoint $x=1$ is absorbing, and no normalized stationary density exists on \((0,1)\).
Finally, if \(cS_{A\cG}=0\), then \(cD_{A\cG}=0\) as well, these reduce to the
selection-free density given above in \eqref{eq:m=1_not12} and \eqref{eq:m=1_12}.

\subsubsection{The full time-dependent solution of the Fokker-Planck equation.} \label{sss:fullsoln}  
Using the stationary density $\pstar$, we obtain the time-dependent solutions of \eqref{eq:fokker-planck}. This is facilitated by formulating the eigenvalue problem for the generator $\cL$, which acts on a test function $f$ via 
\beq \cL f=\frac1{2\pstar}\frac{d}{dx}\left(\sigma^2(x)\pstar f'\right). \label{eq:cL_pstar} \eeq 
The boundary value problem
\beq -\cL f=\lambda f,\quad f'(0)=0=f'(1) \label{eq:cL_slp} \eeq 
is a regular Sturm--Liouville problem with weight function $\pstar$. Its eigenvalues satisfy
\[0=\lambda_0<\lambda_1\leq\ldots\to\infty.\]
Its eigenfunctions can be normalized to constitute an orthonormal basis for the weighted Hilbert space $L^2((0,1),\pstar(x)\,dx)$. Denoting these eigenfunctions by $\varphi_k$, we have that $\varphi_0=1$. Moreover,

\[\cLstar(g)=\frac{d}{dx}\left(\frac{\sigma^2\pstar}2\frac{d}{dx}\left(\frac g{\pstar}\right)\right).\]
From this we see that $g$ is an eigenfunction of $\cLstar$ if and only if $g/\pstar$ is an eigenfunction of $\cL$, with the corresponding boundary conditions. Consequently, the time-dependent solution $\rho(t,x)$ to \eqref{eq:fokker-planck} has the expansion
\begin{equation}
\rho(t,x)=\pstar(x)\left[1+\sum_{n\geq1}c_n e^{-\lambda_n t}\varphi_n(x)\right],
\end{equation}
where the coefficients $c_n$ are determined by the initial condition $\rho(0,x)$. Since the eigenvalues $\lambda_n>0$ for all $n\geq1$, the solution approaches $\pstar$ as $t\to\infty$.  The rate of convergence to the stationary density is exponential and controlled by the smallest positive eigenvalue $\lambda_1$, which can be numerically computed given the values of all variables in Table \ref{tabel_results}. Although the eigenvalues of $-\mathcal L$ cannot generally be found
in closed form, they can be approximated numerically. Under our
assumptions, \eqref{eq:cL_slp} is a regular Sturm--Liouville problem, and it is a classical result \cite{Zettl2005} that the 
eigenvalues satisfy
\[
\lambda_n\sim
\frac{\pi^2 n^2}
{\left(\displaystyle\int_0^1
\sqrt{2/\sigma^2(x)}\,dx\right)^2}
\qquad (n\to\infty).
\]

\section{Discussion}\label{s:disc}
Incorporating variability among individuals connects the internal composition of a species to its abundance dynamics in a finite ecological community. Earlier models have shown that variation among individuals can alter competitive outcomes and coexistence \cite{Lichstein2007Intraspecific,Hart2016IndividualVariation, MendenDeuerRowlett2014, MendenDeuerRowlett2021}.  Like Lichstein et al.~\cite{Lichstein2007Intraspecific} and Hart et
al.~\cite{Hart2016IndividualVariation}, we allow individuals within a species to
differ in competitive performance. These models differ from ours in how
individual-level variation enters species-level dynamics.
Lichstein et al.\ calculate a species' probability of winning
an open site, where the winner is the best performer among
the individuals arriving there. Hart et al.\ incorporate
individual variation in demographic traits and competitive
responses into population growth and interaction strengths.
Here, each local competitive event pairs one individual of the focal species 
$A$ with one individual of the aggregate local community $\cG$. We calculate the
probabilities of a win, loss, or tie from their full
competitive-ability distributions as in \cite{MendenDeuerRowlett2014, MendenDeuerRowlett2019, MendenDeuerRowlett2021, RowlettKarlssonNursultanov2022} and use these probabilities to determine the transition rates of the finite-community model. Thus, differences in distributional shape can affect
both the direction and the stochasticity of local abundance
changes. 

Our results show that, in a finite local community, even the mean and variance of competitive ability distributions do not suffice to predict a focal species' fate: competitive ability distributions with identical mean and variance can produce different extinction and fixation probabilities and, with immigration, different stationary abundance densities. The shape of the distribution of competitive abilities among individuals has also been shown to affect fixation in spatially heterogeneous population-genetic models
\cite{MahdipourShirayeh2017Fixation}.  Here, this distribution shape determines the outcome of individual competitive encounters which in turn is used to calculate the Markov chain.  Its diffusion approximation then provides a way to track the role of the distribution shape from the individual encounters to their consequences for local abundance (Figure~\ref{fig:individual-variability-schematic}).

\subsection{Distribution shape and ecological equivalence}

The distributions of individual competitive abilities determine the
directional advantage $D_{A\cG}$ and the probability of a decisive
encounter $S_{A\cG}$. In our finite-community model, these encounter
parameters govern fixation and, with immigration, stationary
abundance. Phenotypic differences among microorganisms, even among
genetically identical cells, can likewise affect population
performance \cite{Ackermann2015}.

Figures~\ref{fig:individual-variability}--\ref{fig:A6-literature-sensitivity}
show that focal distributions with identical means and variances can
be disadvantaged, neutral, or advantaged against the same aggregate
$\cG$. Thus, competitive success depends on the distributions
\emph{relative to one another}; the examples do not imply that
greater variance is beneficial. Competitive ability is a quantity
defined by the encounter rule, whose relationship to a measured trait
or to fitness must be established for a particular system.

A useful limiting case clarifies when distribution shape matters.
Suppose that $F_A$ and $F_{\cG}$ are independent and continuously
distributed on $[0,1]$. Writing
$H_{\cG}(a)=\Pr(F_{\cG}\leq a)$, we have no ties and therefore
\[
S_{A\cG}=1,\qquad
D_{A\cG}=2\mathbb E[H_{\cG}(F_A)]-1.
\]
If $F_{\cG}$ is uniform, then $H_{\cG}(a)=a$, so
\[
D_{A\cG}=2\mathbb E[F_A]-1.
\]
Against this aggregate, equal focal means yield identical abundance
processes when the remaining parameters are fixed. If $H_{\cG}$ is
quadratic, the first two moments of $F_A$ suffice instead. For other
aggregate distributions, features beyond those moments can affect
encounter outcomes. In our discrete examples, ties may occur and
$S_{A\cG}=1-\Pr(F_A=F_{\cG})$.

Individual variability determines the encounter probabilities,
whereas demographic stochasticity arises from random abundance
changes in the finite community. Because the abundance process
depends on the ability distributions through $D_{A\cG}$ and
$S_{A\cG}$, abundance data alone cannot distinguish distributions
that give the same values of these parameters.

\subsection{Recovery of Hubbell's neutral model}
 The model retains Hubbell's fixed-community-size construction and recovers its neutral local dynamics as a special case \cite{Hubbell2001}. Without immigration, setting $c=1$, $W_{A\cG}=L_{A\cG}=1/2$, and $T_{A\cG}=0$ gives
\begin{equation}
p_n^+=p_n^-=\frac{n(J-n)}{J(J-1)}.
\label{eq:hubbell_transitions_discussion}
\end{equation}
These are the neutral birth--death transition probabilities under sampling without replacement as in \cite{Hubbell2001}.

With immigration included, the same symmetric competition rule recovers neutral local replacement combined with immigration from a fixed metacommunity. Hubbell's neutrality is defined by ecological equivalence in per-capita demographic rates, rather than simply by equality of a measured mean trait. Our results identify how differences in individual distributions can break this equivalence even when their means agree. They also allow heterogeneous individuals to generate neutral focal-species abundance dynamics. 
This connects individual variability with neutral demographic dynamics without equating neutrality with an absence of variation.

\subsection{Fixation, demographic stochasticity, and community size}
In the absence of immigration, the common factor \(x(1-x)\) in the drift and diffusion coefficients cancels in the fixation equation \eqref{eq:ode_trump}.  Fixation therefore depends on the dimensionless number \(\wp=2JD_{A\mathcal G}/S_{A\mathcal G}\), which measures directional competitive advantage relative to demographic randomness.  Solving \eqref{eq:ode_trump} yields the familiar exponential form \eqref{eq:soln_nodrift} from Kimura's theory \cite{Kimura1962} for genic selection. There is a correspondence between governing dimensionless parameters in Kimura's theory and our model
\beq 4 N s \longleftrightarrow \wp = \frac{2J D_{A \cG}}{S_{A \cG}}. \label{eq:kimura_us} \eeq 
Above, $s$ is the selective advantage, and $N$ is the number of diploid individuals.   The parameter $\wp$ plays the role of Kimura's scaled selection parameter $4 N s$:  both quantify directional advantage relative to random fluctuations and determine the shape of the fixation probability curve.  Here, the analogous parameter $\wp$ is determined by the outcomes of encounters between individuals.  Consequently, different shapes of competitive-ability distributions can yield different fixation probabilities
even when their competitive-ability distributions have identical means and variances.

In a closed community, $\wp$ measures directional competitive advantage
relative to demographic stochasticity. As in classical fixation theory
\cite{Kimura1962}, an advantage increases the probability of fixation
without guaranteeing it. Community size therefore matters: smaller
communities are more susceptible to ecological drift, consistent with
earlier metacommunity models \cite{OrrockWatling2010}. In our model,
however, this balance depends on the full distributions of individual
competitive abilities through $D_{A\cG}$ and $S_{A\cG}$; equal means
and variances need not produce equal fixation probabilities. Changing
$c>0$ rescales the pace of competition in a closed community, but
does not change these probabilities or generate a restoring force
toward coexistence.

\subsection{Immigration and long-term local abundance}

Immigration connects local competitive outcomes to the regional availability of species. As in Hubbell's neutral theory, the immigration probability, $m$, is the probability that a local update replaces a resident with an individual drawn from the metacommunity, while $\pi_A$ is the proportion of that regional pool belonging to the focal species $A$ \cite{Hubbell2001}. These quantities describe different aspects of dispersal: how frequently immigrants replace local individuals and which species those immigrants represent. The local aggregate competitor, $\cG$, instead comprises all species other than $A$ already present in the local community. This distinction follows the separation between local interactions and regional dispersal emphasized in metacommunity theory \cite{Leibold2004}. Our extension allows the outcome of local competition to depend on individual competitive-ability distributions, while retaining a specified regional supply of immigrants.

The possibility that immigration sustains local occurrence despite an unfavourable competitive environment connects our results to source--sink theory and the mass-effects perspective on metacommunities \cite{Pulliam1988,Leibold2004}. In source--sink theory, immigration can maintain populations where local reproduction does not compensate for mortality; mass effects describe how dispersal influences local occurrence despite differences in local performance. Here, a species that loses competitive encounters more frequently than it wins can nevertheless return after local extinction, provided that it remains represented in the regional pool. This is a comparable demographic effect, although our model does not explicitly represent different habitat qualities or separate source and sink populations. Continued regional input also does not guarantee presence at every moment. In the finite Markov chain, a species can disappear locally and subsequently be reintroduced. Recurrent local occurrence must therefore be distinguished from persistence in a closed community.

The parameters governing immigration separate regional availability from the strength of local competitive sorting. Increasing the immigration probability gives regional replacement a greater role relative to local competition; when every update is an immigration event, competitive-ability distributions no longer affect abundance. Increasing the regional proportion $\pi_A$ of the focal species $A$ makes arriving individuals more likely to belong to $A$, without changing its success in a local competitive encounter. This separation is shared with neutral immigration models \cite{Hubbell2001}, but here it permits regional supply to oppose a competitive disadvantage arising from distribution shape. Immigration also introduces randomness through the identities of arriving and replaced individuals. Consequently, stronger immigration need not reduce abundance variability in every parameter regime, even though it increases the contribution of regional composition to local dynamics.

The encounter parameter, \(c\), describes a different process: it is the probability that a sampled pair—one individual from the focal species \(A\) and one from aggregate local community of all other species \(\mathcal G\)—competes. Conditional on competition, their competitive ability distributions determine whether \(A\) wins, loses, or ties.
Reducing this probability weakens local competition relative to immigration when immigration is held fixed. This allows a simple comparison with spatial models in which the arrangement of individuals affects their interactions. For example, Bolker and Pacala \cite{BolkerPacala1997} explicitly model spatial covariances, whereas our parameter summarizes encounter frequency without tracking spatial arrangement. It cannot reproduce all consequences of aggregation, dispersal distance, or habitat differentiation. 
Moreover, in a closed community, lowering the encounter probability $c$ slows changes in abundance but leaves fixation probabilities unchanged. It does not cause a species to gain an advantage when rare and therefore does not, by itself, stabilize coexistence \cite{Chesson2000}.

\begin{figure}[p]
\centering
\includegraphics[
    width=\linewidth,
    height=0.82\textheight,
    keepaspectratio
]{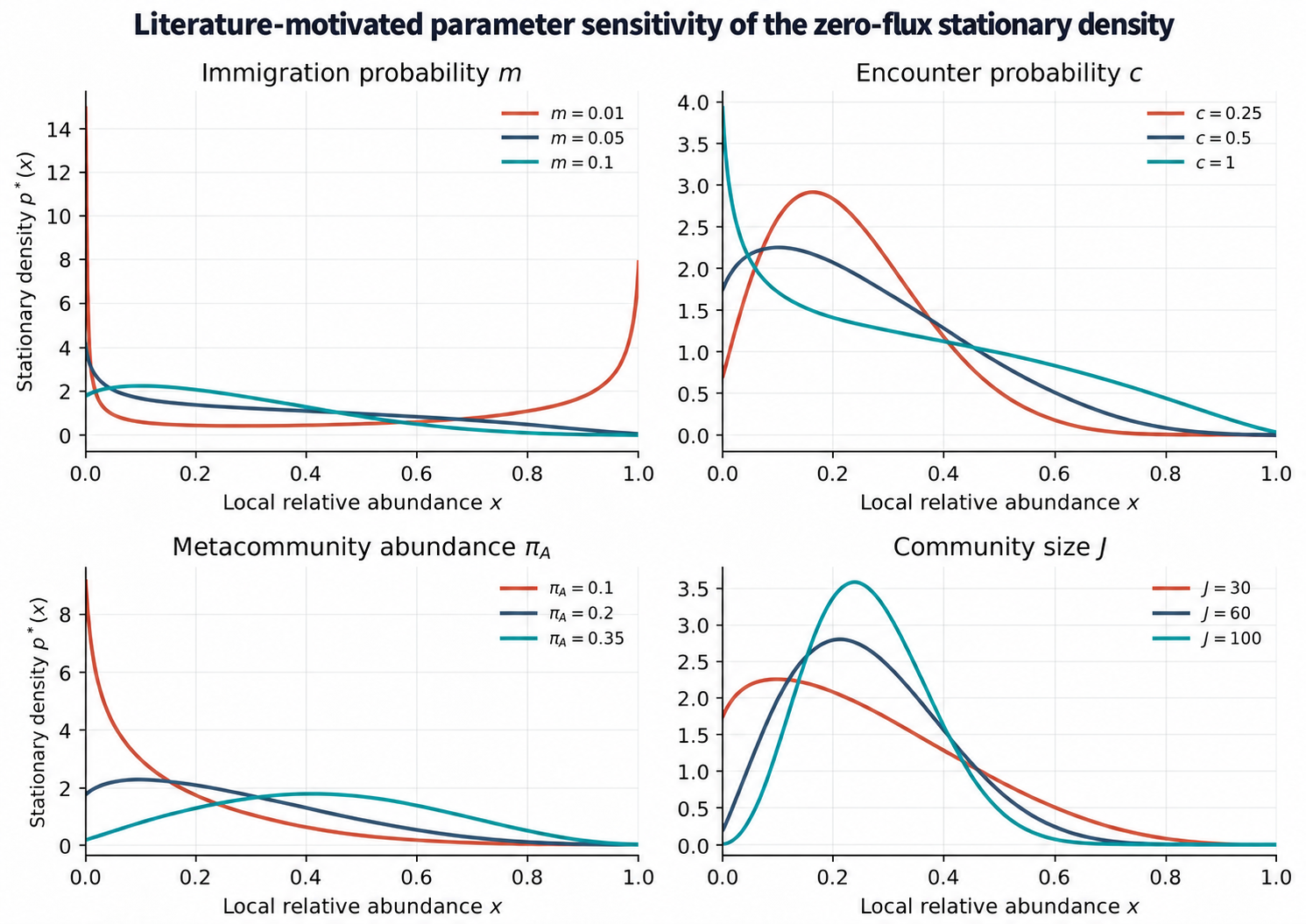}
\caption{Literature-motivated sensitivity of the zero-flux stationary
density for focal species $A_6$ with competitive ability density given in Figure \ref{fig:A4-A6-abilities}. The reference values are
$m=0.10$, $c=0.5$, $\pi_A=0.20$, and $J=30$; only the parameter
named in each panel varies. Marine bacterioplankton experiments
using 1\% and 10\% additions of immigrants motivate the range
of immigration values \cite{BouvierEtAl2012}. Observed abundances
of the SAR11 lineage motivate the scale of $\pi_A$
\cite{BeckerEtAl2019}, while experimental microbial droplets
motivate the small-community values of $J$
\cite{TanSaleskiLin2022}. These studies motivate parameter
scales rather than calibrate this hypothetical $A_6$ community.}
\label{fig:A6-literature-sensitivity}
\end{figure}

Local community size, $J$, is the total number of individuals sharing the fixed local community, and determines the importance of demographic stochasticity: random abundance changes caused by individual replacement events. This role is consistent with the finite-population construction of neutral theory and with stochastic ecological models derived from individual events \cite{Hubbell2001,BlackMcKane2012}. Larger communities experience smaller fluctuations in relative abundance, but their expected composition still reflects both competition and regional input. The stationary density, $\pstar(x)$, describes the long-term probability distribution of the focal species' relative abundance, $x$, rather than a single abundance that every community must attain. Figures~\ref{fig:stationary_density2} and \ref{fig:A4-A6-stationary} show how different competitive-ability distributions produce different stationary abundance distributions under the same immigration regime. Figure~\ref{fig:A6-literature-sensitivity} separates the effects of immigration, encounters, regional composition, and community size within this framework by varying the parameter values based on empirical estimates of these values for marine microbes \cite{BouvierEtAl2012,BeckerEtAl2019, TanSaleskiLin2022}.

These stationary results concern a continuous approximation to the original replacement process. The need to distinguish an individual-based stochastic model from its population-level approximation is emphasized by Black and McKane \cite{BlackMcKane2012}. In our immigration model, zero-flux boundary conditions specify a reflected diffusion: probability cannot flow beyond relative abundances zero and one. This is an additional boundary specification because the interior stochastic differential equation, with nonzero diffusion at the endpoints, does not itself confine trajectories to that interval. The reflected model retains bounded abundance but does not exactly reproduce discrete waiting times at local extinction or fixation. Comparisons with the Markov chain are therefore particularly relevant for small communities or weak immigration. The slowest relaxation time describes how the abundance distribution loses dependence on its initial state under fixed conditions; it should not be interpreted as a general measure of ecological resilience to environmental change.

\subsection{Scope, empirical interpretation, and extensions}

Our framework combines competitive differences, ecological drift, and dispersal, three of the processes organized within Vellend's general community theory \cite{Vellend2016}. Its purpose is to determine how variation among individuals changes demographic outcomes, rather than to establish that such variation necessarily promotes stable coexistence, although variation has been shown to promote coexistence \cite{MendenDeuerRowlett2014, MendenDeuerRowlett2019, MendenDeuerRowlett2021}. Chesson \cite{Chesson2000} distinguishes reductions in average competitive differences from stabilizing mechanisms that favour recovery from low abundance. Likewise, Stump et al.~\cite{Stump2022} show that the consequences of within-species trait variation depend on how traits affect competitive differences and stabilization. Our fixed-distribution model isolates a particular pathway: individual encounter outcomes determine the competitive bias and demographic fluctuations of a focal species. Without immigration or a restoring advantage when rare, the nondegenerate finite process ultimately reaches local extinction or fixation. Its results therefore complement coexistence analyses by quantifying demographic consequences, without establishing a general solution to the maintenance of multispecies diversity.

The species' \emph{strategy} is its probability distribution of individual competitive abilities; it does not imply deliberate behaviour by organisms. Holding this distribution fixed allows us to separate its demographic effects from changes in the traits themselves. Phenotypic heterogeneity among genetically identical microorganisms provides a biological motivation for this representation \cite{Ackermann2015}, but its presence does not by itself justify a constant distribution. Replacement preserves species identity in our model without explicitly tracking the inheritance of an individual's competitive ability. A fixed distribution is therefore most plausible when its shape is approximately maintained or regenerated over the period considered. This assumption differs from models that include heritable trait variation and evolutionary changes, such as that of Barab\'{a}s and D'Andrea \cite{BarabasDAndrea2016}. If successful phenotypes become progressively more common within a species, its competitive-ability distribution must also be allowed to change.

The aggregate competitor is another deliberate simplification. It represents the combined distribution of competitive abilities among all local individuals not belonging to the focal species, rather than a separately defined biological species. Metacommunity theory treats interactions among species and movement among communities as potentially coupled processes \cite{Leibold2004}; our one-dimensional model instead holds the effective local background and regional composition fixed. This makes it possible to follow the focal species while avoiding a full description of every other species' abundance. However, changes in the relative abundances of those other species can change the aggregate distribution even if each constituent species retains its own strategy. The examples therefore compare separate focal-species scenarios against a common background, rather than simultaneous dynamics of all illustrated focal species. Extending the model to a changing community would require tracking its composition or justifying an approximation for the aggregate distribution.

Other assumptions specify which biological mechanisms are represented. The fixed-community-size assumption follows Hubbell's construction of a saturated community, where one individual's gain is accompanied by another's loss \cite{Hubbell2001}. It excludes changes in total abundance arising from changing resource supply. The common bound on mean competitive ability constrains the strategies being compared, as in the preceding distribution-based game model \cite{RowlettKarlssonNursultanov2022}; it is not an independently established physiological law. Trait-based phytoplankton ecology instead relates performance to particular mechanisms, such as nutrient acquisition, light use, and associated trade-offs \cite{LitchmanKlausmeier2008}. A biological application of our model would need to identify how such mechanisms justify an effective competitive-ability scale and its constraints. Competitive ability here measures success under the specified replacement rule and should not automatically be equated with an observed trait or lifetime reproductive fitness.

The empirical relevance of individual-level distributions is supported by Fontana et al.~\cite{FontanaThomasMoldoveanu2018}, who found that individual-level trait diversity predicted phytoplankton community properties better than species richness or evenness. Their findings motivate measurements beyond species-level summaries, but do not directly test the demographic mechanism proposed here. A more specific test would compare replicated communities whose focal populations have similar trait means and variances but different distribution shapes, while controlling the background community and immigration regime. The measured trait would first need to predict replacement success, consistent with the emphasis on trait--performance relationships in trait-based ecology \cite{LitchmanKlausmeier2008}. The win, loss, and tie probabilities are the chances that the focal individual replaces its competitor, is replaced, or produces no abundance change during an effective encounter. Estimating these probabilities would connect individual measurements to predictions for local abundance. Abundance data alone cannot uniquely recover the underlying distributions, because different strategies can yield the same encounter probabilities.

An extension would allow these strategies to evolve while abundance changes stochastically. Barab\'{a}s and D'Andrea \cite{BarabasDAndrea2016} demonstrate that heritability and trait evolution can change conclusions drawn from nonheritable individual variation. Within the distribution-based game framework, Karlsson et al.~\cite{KarlssonGerleeRowlett2025} develop adaptive dynamics, a deterministic description of changes in strategies driven by competitive performance. They identify stationary strategies with game equilibria, but also find nonstationary dynamics and instability. Our analysis addresses a complementary question by holding strategies fixed and following demographic uncertainty in finite communities. Coupling the two descriptions would allow the effects of inheritance, changing distribution shape, and immigration to be studied together. Such a model would need to specify the relative time scales of trait change and abundance change; convergence of abundance toward a stationary distribution does not imply that the competitive strategies themselves converge.

The contribution of the present model is thus an explicit demographic connection between the distribution-based representation of competition and the finite-community processes of neutral theory \cite{RowlettKarlssonNursultanov2022,Hubbell2001}. Populations with identical mean competitive ability and variance can have different competitive outcomes against the same background, and these differences can be expressed in fixation probabilities or long-term abundance distributions, as illustrated in Figures ~\ref{fig:A4-A6-fixation} and ~\ref{fig:A4-A6-stationary}. The relevant prediction is that distribution shape can improve demographic predictions when the trait--competition relationship and the surrounding community are specified. This complements the broader finding that within-species variation can have contrasting ecological effects \cite{Stump2022} by identifying a tractable pathway through which those effects enter local abundance dynamics.

\section*{Acknowledgements}
We are grateful to Susanne Menden-Deuer and Bingzhang Chen for discussions and correspondence that contributed to the development of this work, and for their comments on the first draft of the manuscript.

\providecommand{\etalchar}[1]{$^{#1}$}
\providecommand{\bysame}{\leavevmode\hbox to3em{\hrulefill}\thinspace}
\providecommand{\MR}{\relax\ifhmode\unskip\space\fi MR }
\providecommand{\MRhref}[2]{%
  \href{http://www.ams.org/mathscinet-getitem?mr=#1}{#2}
}
\providecommand{\href}[2]{#2}

    \end{document}